\documentclass[prd,aps,superscriptaddress,twocolumn,%
showpacs,preprintnumbers,amsmath,amssymb,nofootinbib]{revtex4-1}

\usepackage{graphicx}
\usepackage{bm}
\usepackage[colorlinks=true, pdfstartview=FitV, citecolor=blue, urlcolor=blue]{hyperref}
\usepackage{slashed}
\usepackage{array}

\renewcommand{\Im}{\mathop{\mathrm{Im}}}
\renewcommand{\Re}{\mathop{\mathrm{Re}}}
\newcommand{\balpha}{\boldsymbol{\alpha}}
\newcommand{\bphi}{\boldsymbol{\phi}}
\newcommand{\bsigma}{\boldsymbol{\sigma}}
\newcommand{\bz}{\boldsymbol{z}}

\newcommand{\bmu}{\boldsymbol{\mu}}

\newcommand{\calV}{\mathcal{V}}
\newcommand{\Vbos}{\calV_{\text{bos}}}

\newcommand{\VGPY}{\calV_{\text{GPY}}}
\newcommand{\gammadec}{\gamma_{\text{dec}}}
\newcommand{\gammaCP}{\gamma_{\text{CP}}}
\newcommand{\arcsinh}{\mathop{\mathrm{arcsinh}}}
\newcommand{\CP}{\mathcal{CP}}
\newcommand{\perm}{\mathcal{P}_W}
\newcommand{\calO}{\mathcal{O}}
\newcommand{\calN}{\mathcal{N}}
\newcommand{\calZ}{\mathcal{Z}}

\newcommand{\bbZ}{\mathbb{Z}}
\newcommand{\Tdec}{T_{\text{dec}}}
\newcommand{\Tcp}{T_{\text{CP}}}
\newcommand{\phic}{\bphi_\text{c}}

\newcommand{\be}{\begin{equation}}      
\newcommand{\ee}{\end{equation}}      
\newcommand{\bea}{\begin{eqnarray}}      
\newcommand{\eea}{\end{eqnarray}}
\newcommand{\im}{\mathrm{i}}
\newcommand{\tr}{\mathrm{tr}}
\newcommand{\diff}{\mathrm{d}}
\newcommand{\p}{\partial}

\begin{document}

\title{Deconfinement and $\mathcal{CP}$-breaking at $\theta=\pi$ in Yang-Mills theories\\ 
and a novel phase for $\mathrm{SU}(2)$}

\author{Shi Chen}
\email{s.chern@nt.phys.s.u-tokyo.ac.jp}
\affiliation{Department of Physics, The University of Tokyo, 
  7-3-1 Hongo, Bunkyo-ku, Tokyo 113-0033, Japan}

\author{Kenji Fukushima}
\email{fuku@nt.phys.s.u-tokyo.ac.jp}
\affiliation{Department of Physics, The University of Tokyo, 
  7-3-1 Hongo, Bunkyo-ku, Tokyo 113-0033, Japan}

\author{Hiromichi Nishimura}
\email{hnishimura@keio.jp}
\affiliation{Research and Education Center for Natural Sciences, 
  Keio University, Kanagawa 223-8521, Japan}
  \affiliation{Research Center for Nuclear Physics (RCNP), Osaka University, Osaka 567-0047, Japan}

\author{Yuya Tanizaki}
\email{yuya.tanizaki@yukawa.kyoto-u.ac.jp}
\affiliation{Yukawa Institute for Theoretical Physics, Kyoto University, 
  Kyoto 606-8502, Japan}

\preprint{YITP-20-78}

\begin{abstract}
  We discuss the deconfinement and the $\mathcal{CP}$-breaking phase transitions at $\theta=\pi$ in Yang-Mills theories. The 't~Hooft anomaly matching prohibits the confined phase with $\mathcal{CP}$
  symmetry and requires $T_{\mathrm{dec}}(\theta=\pi) \le T_{\mathrm{CP}}$,
  where $T_{\mathrm{dec}}(\theta=\pi)$ and $T_{\mathrm{CP}}$ denote
  the deconfinement and the $\mathcal{CP}$-restoration temperatures, respectively, at $\theta=\pi$. We analytically study these two phase transitions in softly-broken $\mathcal{N}=1$ supersymmetric Yang-Mills theories on small $\mathbb{R}^3\times S^1$ with the periodic boundary condition for gluinos. For most gauge groups except $\mathrm{SU}(2)$ in this model, we find that the inequality is saturated, so deconfinement and $\mathcal{CP}$ restoration occur simultaneously. We demonstrate special features of the $\mathrm{SU}(2)$ gauge theory: There is a finite window of two temperatures, $T_{\mathrm{dec}}(\pi)<T_{\mathrm{CP}}$, which indicates the existence of a novel $\mathcal{CP}$-broken deconfined phase. We also discuss an implication of the novel phase for domain walls and their junctions.
\end{abstract}
\maketitle

\section{Introduction}

Physics of the $\theta$ vacuum is traced back to instanton studies in
Yang-Mills (YM) theories and has become common knowledge in the field of
quantum chromodynamics (QCD)~\cite{Belavin:1975fg, Callan:1976je, Jackiw:1976pf, tHooft:1976rip}. 
Although the QCD $\theta$-angle itself is consistent
with zero in our universe~\cite{Baker:2006ts}, it has been an interesting problem to
quantify the curvature of the $\theta$-dependent vacuum energy (i.e.,
the topological susceptibility~\cite{Witten:1979vv}), and even to reveal
the whole phase structure as a function of nonzero $\theta$ (see, e.g., Refs.~\cite{Witten:1980sp, tHooft:1981bkw, Ohta:1981ai, Cardy:1981qy, Cardy:1981fd, Wiese:1988qz, Affleck:1991tj, Creutz:1995wf, Creutz:2009kx, Smilga:1998dh, Witten:1998uka, Halperin:1998rc, Creutz:2003xu, Boer:2008ct, Aoki:2014moa, Mameda:2014cxa, Verbaarschot:2014upa}).  
Theoretical interest in $\theta\neq 0$ physics is not limited to QCD
but spreads over various research areas such as axion dynamics,
topological phases of matters, chirally induced effects, etc.

A $\theta$ term explicitly breaks $\CP$ symmetry except for either
$\theta=0$ or $\theta=\pi$.  It is, however, known that a first-order phase transition occurs at $\theta=\pi$ because of the spontaneous breakdown of $\mathcal{CP}$, which is sometimes referred to as Dashen's
phenomenon~\cite{Dashen:1970et}.  There must be a nontrivial interplay
between the realization of $\CP$ symmetry at $\theta=\pi$ and
nonperturbative properties of theory, namely, confinement and chiral
symmetry breaking.  In particular, one interesting question is the
fate of spontaneous $\CP$ breaking at high temperature, where the
deconfined phase is achieved.  If the $\CP$-breaking mechanism is
inherent to confinement, as is intuitively anticipated, $\CP$ would be
restored in the deconfined phase.  There are a number of theoretical
efforts but the problem is highly nonperturbative, and to make matters
worse, the Monte-Carlo simulation on the lattice does not work for
non-small $\theta$ due to the sign problem.

A breakthrough has recently been brought by modern development in
quantum field theory based on the 't~Hooft anomaly
matching~\cite{tHooft:1979rat, Frishman:1980dq}. The 't~Hooft anomaly characterizes
an obstruction to gauging a global symmetry.  
Importantly, the 't~Hooft anomaly is renormalization-group invariant, which is called the anomaly matching condition. 
The applicability of anomaly matching has been limited to continuous
chiral symmetry until recently, but deeper understandings
on topological phases have elucidated that it generalizes to any kind of symmetries in quantum field theories~\cite{Wen:2013oza, Kapustin:2014lwa, Cho:2014jfa, Wang:2014pma}.  
As a consequence of anomaly matching, symmetry breaking, massless excitations, and/or topological orders must be realized, and this is a very strong constraint on possible phase
structures.  
Successful applications
include one interesting example that has shed new light on the
mechanism of Dashen's phenomenon~\cite{Gaiotto:2017yup}, which triggers many new applications of anomaly matching to QCD and QCD-like theories~\cite{Tanizaki:2017bam, Kikuchi:2017pcp, Komargodski:2017dmc, Komargodski:2017smk, Shimizu:2017asf, Wang:2017loc,Gaiotto:2017tne, Tanizaki:2017qhf, Tanizaki:2017mtm,  Yamazaki:2017dra, Guo:2017xex,  Sulejmanpasic:2018upi, Tanizaki:2018xto, Yao:2018kel, Kobayashi:2018yuk,  Tanizaki:2018wtg,Anber:2018jdf,   Anber:2018xek, Armoni:2018bga, Yonekura:2019vyz, Nishimura:2019umw, Karasik:2019bxn, Misumi:2019dwq, Cherman:2019hbq}.

Pure $\mathrm{SU}(N)$ YM theory enjoys $\mathbb{Z}_N$ center symmetry, which characterizes confinement/deconfinement phases and is recently formulated as the $1$-form symmetry~\cite{Gaiotto:2014kfa}. 
An important observation in Ref.~\cite{Gaiotto:2017yup} is that there exists a mixed 't~Hooft anomaly between center symmetry and $\CP$ symmetry at $\theta=\pi$. 
Under an assumption that the YM mass gap does not vanish at any values of $\theta$, anomaly matching naturally requires the spontaneous $\CP$ breaking at $\theta=\pi$, which is consistent with the large-$N$ analysis~\cite{Witten:1998uka}. 
Moreover, there is another interesting feature of this mixed anomaly: It survives even at finite temperatures, which means that the confined phase has to break the $\CP$ symmetry spontaneously at $\theta=\pi$ for any temperatures. 
This gives an inequality to phase-transition temperatures, 
\be
\Tdec(\theta=\pi) \le \Tcp,
\ee 
where $\Tdec(\theta=\pi)$ is the deconfinement
temperature at $\theta=\pi$ and $\Tcp$ is the $\CP$-restoration temperature.  

The anomaly constraint is so powerful, and nevertheless, we would
emphasize that model analyses in a concrete shape should be useful for
us to gain an intuitive feeling and look into microscopic details.
This has motivated the present work.  Besides, the constraint is
imposed as an inequality, and an intriguing question is how the
inequality is derived and whether the inequality is saturated or not.
If the spontaneous breakdown of $\CP$ dynamically needs a confining vacuum, the only
possibility would be $\Tdec(\pi) = \Tcp$, and this is what we would naively expect. 
However, if we find a theory with $\Tdec(\pi) < \Tcp$, this signifies a novel phase of deconfined and yet
$\CP$-broken matter.  Moreover, as we will discuss later, if the
latter is the case, the phase structure itself is highly unique.  In
fact, we will see that the $\mathrm{SU}(2)$ YM theory has an exotic phase diagram with a
first-order boundary and a second-order boundary crossing each other.

We are ultimately interested in the pure YM theory and QCD,
but the analytical approach to these theories in the confinement
regime is still far from feasible.  We will, therefore, take an
alternative strategy to deform QCD into some relevant but accessible
theories.  
One natural choice is softly-broken $\mathcal{N}=1$ supersymmetric YM (SYM) theory. 
This is just a pure YM theory coupled to single adjoint Weyl fermion, called gluino, which enjoys supersymmetry (SUSY) when the gluino is massless. 
This theory could turn back into the usual YM
theory when gluino mass is much larger than the strong scale.  
The greatest advantage in
this extension is that $\calN=1$ SYM exhibits
confinement in the weak coupling region by putting the theory on
$\mathbb{R}^3\times S^1$ with the periodic boundary condition for
gluinos, so that we can reliably
investigate the confinement-deconfinement phase transition in a semiclassical manner~\cite{Davies:2000nw, Unsal:2007jx,Unsal:2007vu,Unsal:2008ch, Poppitz:2011wy, Poppitz:2012sw,Poppitz:2012nz, Argyres:2012vv,Argyres:2012ka, Anber:2011gn, Anber:2013sga, Anber:2014lba, Anber:2015wha}.  
This
implies that a small SUSY breaking controls the order counting of the
analytical calculations.  We cannot directly extrapolate our results
toward the pure Yang-Mills theories, but we can at least deduce general
tendencies and gain a hint to speculate a favorable scenario.

Here, we shall summarize what we find in the present work.  We will
analytically evaluate the infrared effective potential in the softly-broken $\calN=1$ SYM theory and quantify the
critical points corresponding to $\Tdec(\theta)$ and $\Tcp$ for various gauge
groups\footnote{Here, we call the inverse size of $S^1$ as the temperature for simplicity, although we are taking the periodic boundary condition for gluinos. Later, we discuss this point more carefully. }.  
In most cases except $\mathrm{SU}(2)$ we observe that deconfinement and
$\CP$ restoration occur simultaneously, i.e. 
$\Tdec(\theta=\pi)=\Tcp$.
In the $\mathrm{SU}(2)$ case, however, we have arrived at a conclusion that the phase
diagram has a window for a novel phase of the deconfined and
$\CP$-broken matter, that is,
\be
\Tdec(\theta=\pi)< \Tcp \quad \text{for} \quad \mathrm{SU}(2). 
\ee 
Since the deconfinement phase transition is
of the second order for the $\mathrm{SU}(2)$ case, the phase diagram has an
intersection between the second-order curve and the first-order
line associated with the $\CP$ breaking at $\theta=\pi$.  As we
mentioned above, to the best of our knowledge, our work is the first
concrete demonstration of such an exotic phase structure.  

Moreover, we will discuss the properties of domain walls and domain-wall junctions for this novel deconfined, $\mathcal{CP}$-broken phase. 
In the confined phase, the $\mathcal{CP}$ domain wall at $\theta=\pi$ supports a topological field theory, and the deconfinement takes place on the wall. 
In the novel phase, however, Wilson lines are not deconfined on the $\mathcal{CP}$ domain wall. 
We show that certain domain-wall junctions must have a nonzero electric charge, so they can be present only with the insertion of external Wilson lines. 

This paper is organized as follows: New results of this paper are presented in sections~\ref{sec:other}-\ref{sec:junction}, and sections~\ref{sec:symmetries}-\ref{sec:model} are devoted to reviews of previous studies to make this paper self-contained.  
In Sec.~\ref{sec:symmetries}, we 
will briefly explain the discrete symmetries relevant to our later
discussions, namely, $\CP$ symmetry and center symmetries.  In
Sec.~\ref{sec:pureYM}, we will make a concise review of the 't~Hooft
anomaly as well as its implication to constrain the phase structure
in the thermal pure YM theory.  There, we will see that an
inequality should hold for deconfinement and $\CP$ restoration.  In
Sec.~\ref{sec:model} we will explain the softly-broken $\mathcal{N}=1$ SYM
theory and illustrate the symmetry
and anomaly features of this model. 
We also review the infrared effective potential of the softly broken $\calN=1$
SYM theory on small $\mathbb{R}^3\times S^1$.  
In Sec.~\ref{sec:other}, we numerically study the phase transitions to show that the inequality is
saturated for most gauge groups except the $\mathrm{SU}(2)$ case. 
In Sec.~\ref{sec:SU(2)}, we will proceed to the analytical evaluation of the
phase structure for the $\mathrm{SU}(2)$ case.
We will uncover the existence of a novel phase
of deconfined and $\CP$-broken matter.  In Sec.~\ref{sec:junction},
we will illuminate a novelty of this deconfined and $\CP$-broken
matter by focusing on 1-form center symmetry and the domain wall
structures there.  Finally, Sec.~\ref{sec:summary} is devoted to the
conclusions.
In Appendix~\ref{app:nonSU}, we extend our discussion for non-$\mathrm{SU}(N)$ gauge groups. 
In Appendix~\ref{app:largeN}, we perform the analytic computation of the deconfinement transition in the large-$N$ limit for the $\mathrm{SU}(N)$ gauge groups. 

\section{Discrete symmetries for deconfinement and $\CP$ breaking}
\label{sec:symmetries}

We will consider pure gauge theories with a gauge group
$G$. Throughout this work, we concentrate on the cases that
$G$ is a simple, simply-connected, and compact Lie group. 
Since the most important result of this paper appears for
$G=\mathrm{SU}(2)$, the reader may safely
regard it as $G=\mathrm{SU}(N)$. 
The periodicity of the $\theta$ angle is $2\pi$, and no discrete $\theta$ angle appears since $G$ is simply connected.

The theory has the center symmetry, $\calZ(G)$. For example, $\calZ(\mathrm{SU}(N))=\bbZ_N$.
Although this is not a symmetry acting on a local operator, it acts on the fundamental Wilson loop $W(C)$ as 
\be
W(C)\to z\cdot W(C),
\ee 
for $z\in \calZ(G)$.\footnote{When $G=\mathrm{Spin}(N)$ with $N=4k$, there are two inequivalent spinor representations.  Then we need to consider two different ``fundamental'' Wilson loops. This corresponds to the fact that $\mathcal{Z}(\mathrm{Spin}(4k))=\mathbb{Z}_2\times \mathbb{Z}_2$. } 
Here, the Wilson loop is given by 
$W(C)=\mathrm{tr}\left[\mathcal{P}\exp\left(\im \oint_C a\right)\right]$, 
where $a$ is the dynamical $G$ gauge field, $C$ is a closed loop, and the trace is taken over the defining representation. 
In a recent terminology, this is called the $\mathcal{Z}(G)$ $1$-form symmetry, denoted as $\mathcal{Z}(G)^{[1]}$, indicating that it does not act on any local operators but acts only on line operators~\cite{Gaiotto:2014kfa}. 

Let us add a technical comment on the fact that a typical order operator for center symmetry breaking is the fundamental Wilson loop. The Wilson loop is a member of the family of Wilson-'t~Hooft loops~\cite{Aharony:2013hda,Kapustin:2005py}. 
Since we are considering the pure YM theory, all the elementary particles belong to the adjoint representation. 
Still, when we declare that the gauge group is $G$, we sum up $G$-bundles in the path integral, rather than $G/\mathcal{Z}(G)$-bundles that cannot be lifted to $G$-bundles.
Correspondingly, the fundamental Wilson lines are well defined as genuine line operators, but the fundamental 't~Hooft lines depend on the topology of a surface attaching to them~\cite{Aharony:2013hda}, so such magnetic lines are not order parameters. 

When we put our system on Euclidean $\mathbb{R}^3\times S^1$ (like a finite-$T$ field theory where the inverse temperature $1/T$ is the $S^1$ period), 
we obtain an important local order parameter, the Polyakov loop. 
The Polyakov loop is the Wilson loop wrapping on $S^1$, and it is a point-like object from the point of view of the $3$D theory. 
Correspondingly, the $1$-from center symmetry in $4$D splits into the 1-form and 0-form parts in $3$D viewpoints:
\be
\Bigl(\mathcal{Z}(G)^{[1]}\Bigr)_{4\mathrm{D}} \rightarrow \Bigl(\mathcal{Z}(G)^{[1]}\Bigr)_{3\mathrm{D}}\times \Bigl(\mathcal{Z}(G)^{[0]}\Bigr)_{3\mathrm{D}}. 
\ee
The $1$-form part acts on spatial Wilson loops, i.e., contractible Wilson loops, while the $0$-form center symmetry acts on the Polyakov loops. 

The 0-form piece of center symmetry, $(\mathcal{Z}(G)^{[0]})_{3\mathrm{D}}$, is important for the deconfinement phase transition at finite $T$. In the low-$T$ confined phase, the 0-form center symmetry is unbroken and the expectation value of the Polyakov loop vanishes. As $T$ goes up, the deconfinement phase transition would occur at $T=\Tdec$. In the high-$T$ deconfined phase the 0-form center symmetry is spontaneously broken and the Polyakov loop acquires a nonzero expectation value. It is known that the 1-form piece of center symmetry does not break even at high $T$ and the spatial Wilson loops show area-law behavior throughout.

In contrast to center symmetry, the theory enjoys (0-form) $\CP$ symmetry only at special values of the $\theta$ angle. In fact, the $\CP$ operation reverses the sign of $\theta$, hence $\CP$ symmetry only occurs at either $\theta=0$ or $\theta=\pi$. The former case is obvious, while the latter is based on the $2\pi$ periodicity of $\theta$, namely, $\theta=-\pi$ returns to $\pi$ by the identification $\theta\sim \theta+2\pi$. 
It is known as Dashen's phenomenon that $\CP$ symmetry is spontaneously broken in the confined phase at $\theta=\pi$, resulting in a first-order phase transition with respect to $\theta$ ~\cite{Dashen:1970et} (see also \cite{Witten:1980sp, tHooft:1981bkw, Ohta:1981ai, Cardy:1981qy, Cardy:1981fd, Wiese:1988qz, Affleck:1991tj, Creutz:1995wf, Creutz:2009kx, Smilga:1998dh, Witten:1998uka, Halperin:1998rc, Creutz:2003xu, Boer:2008ct, Aoki:2014moa, Mameda:2014cxa, Verbaarschot:2014upa}). The interplay between confinement and Dashen's phenomenon has been a longstanding problem, and a modern theoretical approach based on the 't~Hooft anomaly provides us with an important clue to clarify it~\cite{Gaiotto:2017yup}.

\section{'t~Hooft anomaly at $\theta=\pi$ and phase structures}
\label{sec:pureYM}

We here give a brief summary of the mixed anomaly of pure $\mathrm{SU}(N)$ YM theory at $\theta=\pi$~\cite{Gaiotto:2017yup}. 
We shall also comment on the same anomaly for other simple, simply-connected, and compact gauge groups $G$ in Appendix~\ref{app:nonSU}. 

Let us begin with a general remark on the modern understanding of 't~Hooft anomaly matching. 
When we have a $d$-dimensional quantum field theory with a global symmetry, we can consider its partition function under the presence of background gauge fields $A$ for the global symmetry, i.e., $Z[A]$. 
When we perform the background gauge transformation, $A\to A+\delta_\xi A$, the partition function is not necessarily gauge invariant. 
Indeed, we often encounter such a situation that the partition function acquires the anomalous phase:
\be
Z[A+\delta_\xi A]=Z[A]\exp\left(\im\int \alpha(\xi,A)\right). 
\ee 
Here, $\alpha(\xi,A)$ is a $d$-dimensional local functional of the background gauge field $A$ and its gauge parameter $\xi$. 
If this anomalous phase cannot be eliminated by adding any $d$-dimensional local counterterms, it is called an 't~Hooft anomaly. 
An important theorem, called anomaly matching, states that the 't~Hooft anomaly does not change under the renormalization-group flow, or more strongly, under any continuous deformations of the theory by local and symmetric Hamiltonians. 
This theorem provides a useful constraint to study strongly-coupled systems nonperturbatively. 
In particular, 't~Hooft anomaly matching excludes a trivially gapped non-degenerate vacuum. This is so because such a vacuum, if manifested, would provide no infrared degrees of freedom to match the anomaly in an infrared effective theory. 

If the system has an 't~Hooft anomaly for continuous symmetry, we can prove that the mass gap has to vanish~\cite{tHooft:1979rat, Frishman:1980dq}. 
When the continuous symmetry is unbroken, the system needs to have color-singlet massless fermions, or to show the conformal behavior at low energies. 
If the continuous symmetry is spontaneously broken to its anomaly-free subgroup, there are massless Nambu-Goldstone (NG) bosons, and the Wess-Zumino-Witten term can match the anomaly~\cite{Wess:1971yu,Witten:1983tw}. 
For discrete symmetries, on the other hand, massless particles do not necessarily appear in order to match the anomaly. 
If the discrete symmetry is spontaneously broken, such a system is typically gapped, but there are degenerate vacua. 
In this case, there are gapless or topological excitations localized on domain walls as in Jackiw-Rebbi mechanism~\cite{Jackiw:1975fn}, and this is crucial to match the anomaly~\cite{Komargodski:2017dmc, Komargodski:2017smk}. 
If the symmetry is unbroken, the anomaly should be reproduced either by a massless excitation or in a form of topological field theory~\cite{Wen:2013oza, Kapustin:2014lwa, Cho:2014jfa, Wang:2014pma}.

Let us now discuss how this technique can constrain the phase diagram of the pure $\mathrm{SU}(N)$ YM theory. 
As we have discussed in Sec.~\ref{sec:symmetries}, the theory has the center symmetry, $\mathbb{Z}_N^{[1]}$, and the $\mathcal{CP}$ symmetry at $\theta=0$ and $\theta=\pi$. 
We shall argue that these two symmetries at $\theta=\pi$ cannot be gauged simultaneously, which shows the presence of the mixed 't~Hooft anomaly. 
For this purpose, following Ref.~\cite{Gaiotto:2017yup}, we introduce the background gauge field for $\mathbb{Z}_N^{[1]}$ and then observe the anomalous violation of $\mathcal{CP}$.

Let us introduce a background gauge field for the center symmetry $\mathbb{Z}_N^{[1]}$, in the form of a $2$-form gauge field $B$~\cite{Kapustin:2014gua,Gaiotto:2014kfa}. 
When our spacetime is the hypertorus $M=T^4$, the introduction of $B$ is equivalent to take the 't~Hooft twisted boundary condition given in Ref.~\cite{tHooft:1979rtg}, and it is mathematically characterized by an element of $H^2(M,\mathbb{Z}_N)$. 
An important effect of $B$ can be found for the instanton number,
\be
Q={1\over 8\pi^2}\int \tr(F\wedge F), 
\ee
where $F=\diff a + \im a\wedge a$ is the $\mathrm{SU}(N)$ field strength. 
When $B$ is absent, the topological charge is quantized as $Q\in \mathbb{Z}$, and there exists a configuration with $Q=1$. This is why we have the periodic $\theta$ angle, i.e., $\theta\sim \theta+2\pi$. 
In the presence of $B$, however, $Q$ has to be redefined so that it becomes invariant under the 1-form gauge transformations, and then it is no longer quantized to integers. 
Indeed, the instanton number $Q$ acquires the $1/N$ fractional piece as~\cite{vanBaal:1982ag} 
\be
Q[B]={N\over 8\pi^2}\int B\wedge B\quad \bmod 1. 
\ee
The right-hand side of the above expression is quantized in a unit of $1/N$. 

Because of the fractionalization of $Q$, we no longer have the $2\pi$ periodic $\theta$ angle, and the partition functions at $\theta$ and $\theta+2\pi$ differ as 
\be\label{eq:anomaly}
Z_{\theta+2\pi}[B] = \exp\left(-\im{N\over 4\pi}\int B\wedge B\right) Z_{\theta}[B]. 
\ee
An extra factor appears, which concludes the mixed 't~Hooft anomaly at $\theta=\pi$~\cite{Gaiotto:2017yup}: By $\mathcal{CP}$ transformation, we see,
\be
Z_{\pi}[B] \mapsto Z_{-\pi}[B]=\exp\left(\im{N\over 4\pi}\int B\wedge B\right) Z_{\pi}[B]. 
\ee
Strictly speaking, we need to examine all possible $4$D local counterterms
to judge if this is a genuine anomaly.
For even $N$, on the one hand, there is no local counterterm that cancels this anomaly, so we get the mixed anomaly between $\mathbb{Z}_{N}^{[1]}$ and $\mathcal{CP}$ at $\theta=\pi$. For odd $N$, on the other hand, a local counterterm cancels the anomaly. Even in the latter case, comparing local counterterms for $\theta=0$ and $\theta=\pi$, we can find that it is impossible to remove the mixed center-$\mathcal{CP}$ anomalies simultaneously for different $\theta$'s~\cite{Gaiotto:2017yup, Tanizaki:2017bam, Kikuchi:2017pcp}. 
This situation is sometimes referred to as a global inconsistency~\cite{Kikuchi:2017pcp, Karasik:2019bxn}, which can be also regarded as a mixed 't~Hooft anomaly in a generalized sense~\cite{Cordova:2019jnf,Cordova:2019uob}.
For both even and odd $N$, this anomaly can be matched by spontaneous breaking of $\mathcal{CP}$ at $\theta=\pi$, and we assume this scenario throughout this paper. 

\begin{figure}
  \includegraphics[width=0.9\columnwidth]{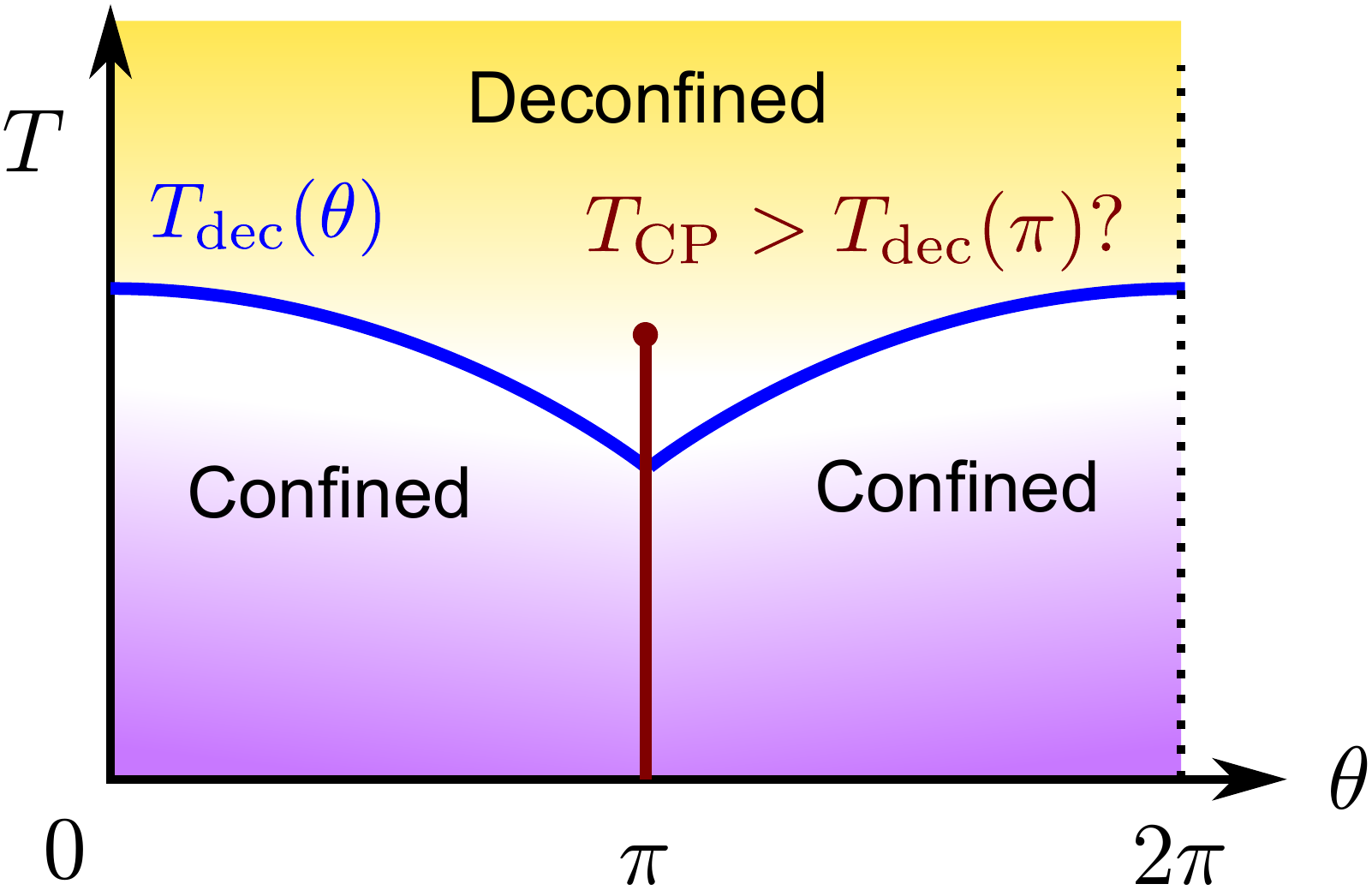}
  \caption{Schematic illustration of a possible phase structure on the $\theta$-$T$ plane.  The blue curves represent the deconfinement transition temperature, $\Tdec$, as a function of $\theta$, and the brown vertical line represents the $\CP$ breaking transition that terminates at $\Tcp$.}
  \label{fig:schematic}
\end{figure}

At finite temperatures, as we already mentioned, center symmetry splits into 1-form and 0-form parts from the 3D perspective, so that the mixed 't~Hooft anomaly above is actually among 1-form center symmetry, 0-form center symmetry, and $\CP$ symmetry. With this anomaly at hand, we can put a strong constraint on the phase structures of the gauge theories on the $\theta$-$T$ plane~\cite{Gaiotto:2017yup}. The most likely scenario at $\theta=\pi$ allowed by the 't~Hooft anomaly is supposed to exclude simultaneous manifestation of $\CP$ and 0-form center symmetries as well as to keep the integrity of 1-form center symmetry. Thus, $\CP$ symmetry should be broken in the 0-form center symmetric phase. That is to say, the deconfinement transition temperature, $\Tdec(\theta=\pi)$, cannot be higher than the $\CP$ restoration temperature, $\Tcp$:
\begin{equation}\label{eq:Tanomaly}
   \Tdec(\theta=\pi) \le \Tcp \,.
\end{equation}
We schematically illustrate this scenario in Fig.~\ref{fig:schematic}. Here, we also note that one should keep in mind other possible but less likely scenarios such as spontaneous 1-form center symmetry breaking and unbroken symmetry with massless excitations or with topological field theories. Hence, it would be important to adopt a dynamical model and investigate which scenario is favored and how it transpires, which motivates us for this work.

Generally speaking, $\Tdec$ may have some $\theta$ dependence with the $2\pi$ periodicity as shown by the blue curves in Fig.~\ref{fig:schematic}. In the case of $G=\mathrm{SU}(2)$, the deconfinement transition is suggested to be of the second order, while it is observed to be a first-order transition for many other gauge groups. The $\CP$ breaking is always accompanied by a first-order phase transition, as shown by the brown vertical line at $\theta=\pi$, which terminates at $T=\Tcp$.

One particularly interesting question is whether the inequality~\eqref{eq:Tanomaly} is saturated, i.e., $\Tdec(\pi)=\Tcp$, or not. If the deconfinement phase transition is of the strong first order, it would be naturally conceivable that the $\CP$ first-order line is chopped off and $\Tdec(\pi)=\Tcp$ would be then derived. If the deconfinement transition is of second order, $\Tdec(\pi)=\Tcp$ would suggest that both symmetries are manifested at the transition point, which is not the only natural way now. A non-trivial temperature window, $\Tdec(\pi) < T < \Tcp$, might exist. If such a temperature window is found, a novel phase of deconfined and $\CP$-broken matter would realize there. This is an extremely interesting possibility; such a crossing shape of two phase boundaries as sketched in Fig.~\ref{fig:schematic} is unique in this system only and, to the best of our knowledge, no similar structure can be found in any other systems. In the subsequent sections, we will discuss such a possibility of the novel phase within the concrete framework of softly-broken $\mathcal{N}=1$ SYM theory.

\section{$\mathcal{N}=1$ supersymmetric YM  theories and the continuity conjecture}
\label{sec:model}

In this section, we explain the softly-broken $\mathcal{N}=1$ SYM theory on $\mathbb{R}^3\times S^1$, and review the adiabatic continuity conjecture between its quantum phase transitions and thermal phase transitions of pure bosonic YM theory. 
We then explain the low-energy effective action on $\mathbb{R}^3\times S^1$ for the $\mathrm{SU}(N)$ gauge group.

\subsection{Supersymmetric YM theories}

The SYM theory of our interest is the Yang-Mills theory with single adjoint Weyl fermion, $\lambda$, which is called gluino:
\be
\begin{split}
    S=&{1\over g^2}\int \tr\left(F \wedge \ast F\right) - {\im\,\theta\over 8\pi^2} \int \tr\left(F\wedge F\right)\\
    &+{2\,\im\over g^2} \int \diff^4 x \,\, \tr\left(\overline{\lambda} \overline{\sigma}^\mu D_\mu \lambda\right), 
\end{split}
\ee
where $D_\mu \lambda=\partial_\mu \lambda+\im[a_\mu,\lambda]$. This theory enjoys $\mathcal{N}=1$ SUSY{}. 
To persist
SUSY, we impose the periodic boundary condition for all the particles
including the gluino along $S^1$.  We note that the finite-$T$ theory
should be anti-periodic for the gluino, and thus we will use $L$ to
denote the $S^1$ periodicity instead of $1/T$.  
This supersymmetric partition function is called Witten index, and, importantly, this theory shows confinement at any $L$ because of the topological property of the index~\cite{Witten:1982df,Witten:1982im}. 
Moreover, this periodic theory has a close
relationship with the thermal pure Yang-Mills theory in a particular
limit as explained below~\cite{Davies:2000nw, Unsal:2007jx,Unsal:2007vu,Unsal:2008ch, Poppitz:2011wy, Poppitz:2012sw,Poppitz:2012nz, Argyres:2012vv,Argyres:2012ka, Anber:2011gn, Anber:2013sga, Anber:2014lba, Anber:2015wha}.

Let us add a mass term to gluino, which breaks SUSY: 
\be
{1\over g^2}\Bigl(m\, \tr(\lambda\lambda)+\mathrm{c.c.}\Bigr). 
\ee 
Assume that the gluino mass is small, $m\ll\Lambda$, to break SUSY softly,
where $\Lambda$ is the dynamical scale of the SYM theory and is
defined through the perturbative running coupling $\alpha_s(\mu)=g(\mu)^{2}/4\pi$ with the renormalization scale $\mu$. 
The two-loop definition of the strong scale is 
\begin{equation}
  \Lambda^3 \equiv \mu^3 \frac{4\pi}{3c_2\alpha_s(\mu)} \,
  \exp\left({-\frac{2\pi}{c_2\alpha_s(\mu)}}\right)\,.
\end{equation}
Here $c_2$ is the dual Coxeter number of $\mathrm{Lie}(G)$, which is $N$ for $G=\mathrm{SU}(N)$.  

\subsection{Adiabatic continuity conjecture}

Now, we point out three immediate
connections between the present model and the thermal pure YM
theory.  Firstly, in the $m\to\infty$ limit, the gluino decouples and
the model reduces to the thermal pure YM theory.  Secondly,
the model shares the fundamental symmetries that motivated this work,
namely, center symmetry and $\CP$ symmetry (at $\theta=0$ or $\pi$).  The gluino belongs to
the adjoint representation, so that center symmetry survives even with
matter fields.  Moreover, the gluino action directly manifests $\CP$
symmetry at $\theta=0$ and $\theta=\pi$.  Thirdly, the mixed 't~Hooft anomaly at $\theta=\pi$ applies to the
present model as well since the gluino in the adjoint representation
does not affect the gauging procedure as described in
Sec.~\ref{sec:pureYM}.

The present model has a theoretical advantage over the thermal pure
YM theory.  This model is much more tractable particularly
about the confinement-deconfinement phase transition, which usually occurs in the strongly-coupled regime.  
For the softly-broken $\calN=1$ SYM, it is found that the confinement-deconfinement transition occurs in the weakly coupled region at $L\ll\Lambda^{-1}$~\cite{Poppitz:2012sw,Poppitz:2012nz,Anber:2014lba}, not in the strong-coupling region. 
Because of its weak-coupling nature, both perturbative and semiclassical calculations are feasible and reliable.  By the virtue of SUSY, as we will discuss later, perturbative contributions are more suppressed than semiclassical ones.  We can hence unravel the phase structure solely by semiclassical computations.

Such semiclassical computations have shown that the $\calN=1$ SYM theory at $m=0$ is a confining theory for any
$L$~\cite{Davies:2000nw}.  For $m\neq 0$ a phase
transition is located at a critical
$L\propto m^{\frac{1}{2}}$ below which deconfinement occurs~\cite{Poppitz:2012sw,Poppitz:2012nz,Anber:2014lba}.
This suggests a natural choice of dimensionless variable:
\be
\gamma\propto {m\over L^2\Lambda^3}.
\label{eq:def_gamma_rough}
\ee  
Then, $\gamma$ plays the role of
a dimensionless temperature in this model and the deconfinement phase
transition takes place at $\gamma=\gammadec$.  We will give the
precise definition of $\gamma$ later in Eq.~\eqref{eq:gamma}. In consistency with the pure Yang-Mills theory, the deconfinement transition in the present model is found to be of the second order for $G=\mathrm{SU}(2)$~\cite{Poppitz:2012sw} and of the first order for other groups~\cite{Poppitz:2012nz,Anber:2014lba}. We also confirm $\theta$ dependent $\gammadec$, i.e, $\gammadec(\theta)$~\cite{Anber:2014lba,Anber:2013sga}. We will go into more details on these behaviors in later discussions.

\begin{figure}
  \includegraphics[width=0.9\columnwidth]{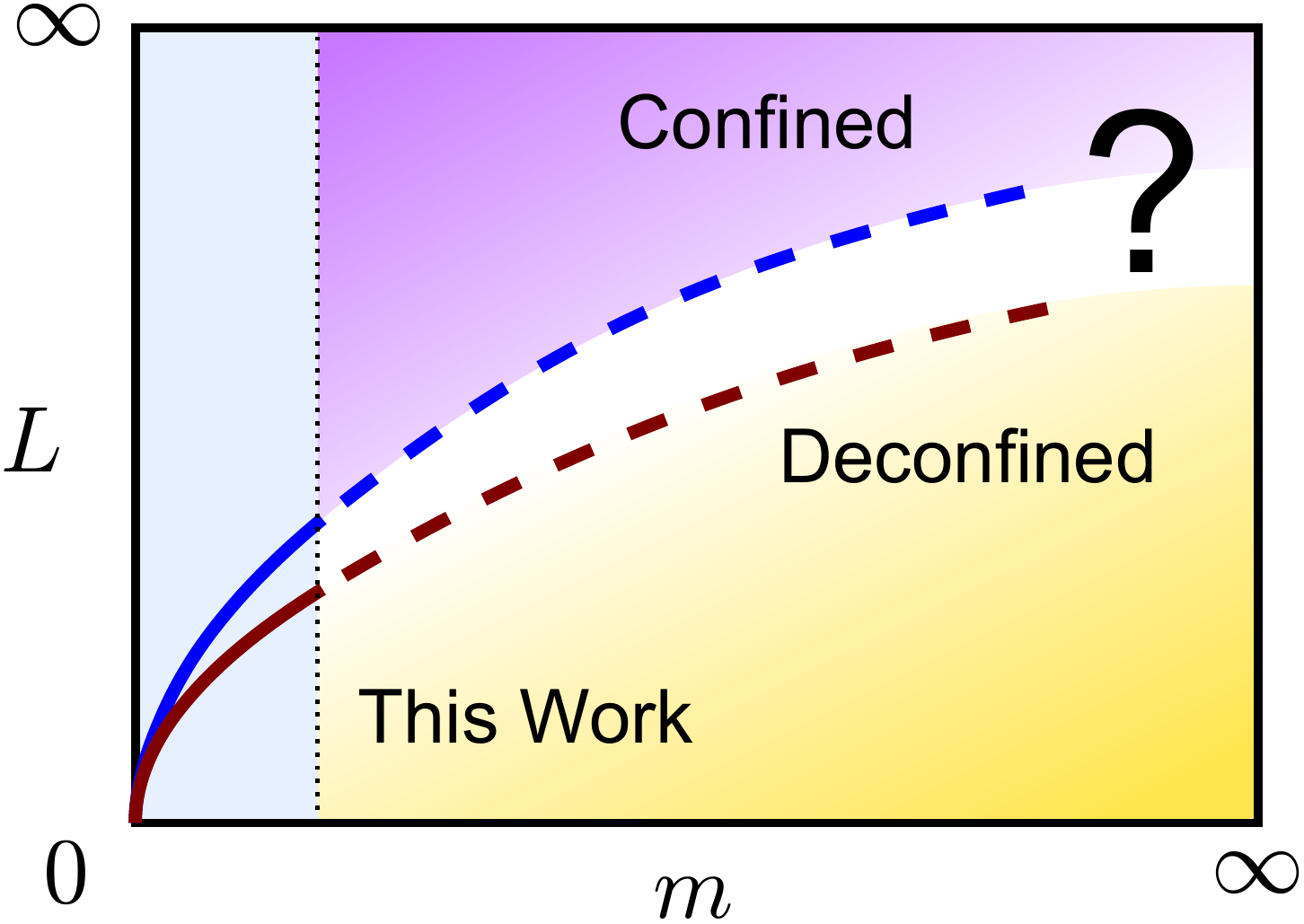}
  \caption{Schematic illustration of a possible phase structure on the $m$-$L$ plane at $\theta=\pi$.  The upper blue line represents the deconfinement transition and the lower brown line represents the $\CP$ restoration transition.  This work covers a small-$m$ region, which is the left corner as indicated by the light-blue shadow.}
  \label{fig:schematic_mL}
\end{figure}

It is then a reasonable
conjecture that this transition line should continue from the
small-$m$ region and end at the deconfinement point in the pure
Yang-Mills theory at $m=\infty$, as indicated by the upper dashed curve
in Fig.~\ref{fig:schematic_mL}.  
That is, the quantum phase transitions of softly-broken SYM on small $\mathbb{R}^3\times S^1$ is smoothly connected to the thermal phase transitions of pure YM theory. 
This is the adiabatic continuity conjecture, and a lot of circumstantial evidences have been obtained so far~\cite{Unsal:2007jx,Unsal:2007vu,Unsal:2008ch, Poppitz:2011wy, Poppitz:2012sw,Poppitz:2012nz, Argyres:2012vv,Argyres:2012ka, Anber:2011gn, Anber:2013sga, Anber:2014lba, Anber:2015wha, Dunne:2012ae, Dunne:2016nmc, Sulejmanpasic:2016llc, Cherman:2016hcd,Cherman:2017tey, Dunne:2018hog, Hongo:2018rpy,  Fujimori:2018kqp, Tanizaki:2019rbk}, although we do not have a direct proof yet. 

In this work, we shall pay special attention to the phase structure
at $\theta=\pi$ in this softly broken SYM theory.
Since this model has a confinement-deconfinement phase transition at
$\gamma=\gammadec$, the mixed 't~Hooft anomaly constrains the possible
realization of center and $\CP$ symmetries in a way as discussed in
Sec.~\ref{sec:pureYM}.  That is to say, $\CP$ symmetry is supposed
to be broken in the confined phase where center symmetry is
manifested.  Therefore, the critical dimensionless ``temperature'',
$\gammaCP$, for $\CP$-restoration should be not smaller than the
deconfinement ``temperature''.  Namely, the following inequality must be
satisfied;
\begin{equation}\label{eq:Ganomaly}
  \gammadec(\theta=\pi) \le \gammaCP \,,
\end{equation}
in this model. In the
thermal pure Yang-Mills theory, this inequality corresponds to Eq.~\eqref{eq:Tanomaly}.  

Figure~\ref{fig:schematic_mL} depicts a scenario
with a finite window between deconfinement and $\CP$ restoration, for
which the critical $L$ for $\CP$ restoration should be smaller than
that for deconfinement.  In what follows below, we will clarify the
possibility of this scenario by calculating the infrared effective
potential.  We will see that for $G=\mathrm{SU}(2)$ the model turns out to have such a window
of $\gammadec(\pi) < \gamma < \gammaCP$, and a novel phase of
deconfined and $\CP$-broken matter emerges.

\subsection{Infrared effective potential}
\label{sec:potential}

Let us now review the low-energy effective action~\cite{Davies:2000nw, Unsal:2007jx,Unsal:2007vu,Unsal:2008ch, Poppitz:2011wy, Poppitz:2012sw,Poppitz:2012nz, Argyres:2012vv,Argyres:2012ka, Anber:2011gn, Anber:2013sga, Anber:2014lba, Anber:2015wha} of the $\calN=1$ SYM theory on small $\mathbb{R}^3\times S^1$. 
For simplicity of notation, we only consider $G=\mathrm{SU}(N)$ in this section. The effective potential for all the simply-connected gauge groups is described in Ref.~\cite{Anber:2014lba}. 

We first discuss the bosonic low-energy degrees of freedom, which are massless at the classical level. 
To find them out, we take the Polyakov gauge, in which $a_0$ is diagonal and independent of $\tau\in S^1$. 
We thus obtain $(N-1)$-component scalar fields, denoted by $\bm{\phi}$. 
Under gauge transformations to the Polyakov gauge, we occasionally encounter the monopole singularities, which will be taken into account soon later and play a very important role. 

For the pure $\mathrm{SU}(N)$ YM theory, $\bm{\phi}$ is classically massless, but acquires the mass of order $g/L$ at the one-loop level~\cite{Gross:1980br}, and its potential is commonly called the Gross-Pisarski-Yaffe (GPY) potential. For the SYM theory, however, SUSY protection is strong enough to prohibit the mass generation of $\bm{\phi}$ at any order of perturbation theory. 
Therefore, as long as the gluino mass $m$ is small enough, we can regard $\bm{\phi}$ as massless quanta at the perturbative level. 
We shall revisit the SUSY-breaking effect on the GPY potential in the last part of this section. 

Next, let us consider the $3$D vector bosons, $\bm{a}$, which are the spatial components of the $\mathrm{SU}(N)$ gauge field. For generic values of $\bm{\phi}$, the $4$D Yang-Mills action produces the mass term, $\tr (F_{0i}^2)=\tr ( [\bm{\phi},\bm{a}_i]^2)$, and the off-diagonal gluons acquire a mass typically of the order of $1/(NL)$. 
The remnant massless degrees of freedom are the $3$D $U(1)^{N-1}$ gauge field, which are the diagonal components of $\bm{a}$. By taking the $3$D Abelian duality, they can be mapped to $(N-1)$-component scalars denoted as $\bm{\sigma}$.

Because of Abelianization, we can study the confinement of the $\calN=1$ SYM theory on small $\mathbb{R}^3\times S^1$ by the weak-coupling semiclassical analysis~\cite{Davies:2000nw, Unsal:2007jx}. 
As $\bm{\phi}$ and $\bm{\sigma}$ come out of gauge fields, they are periodic scalar fields, with the kinetic term;
\be\label{eq:kinetic}
{1\over g^2 L}\Bigl|\diff \bphi\Bigr|^2 + {g^2\over 16\pi^2 L}\Bigl|{\diff \bsigma+{\theta\over 2\pi}\diff \bphi}\Bigr|^2.
\ee
To see the periodicity, it is convenient to express $\bm{\phi}$ and $\bm{\sigma}$ as
\be
\bm{\phi}=\sum_{n=1}^{N-1}\phi_n \bm{\alpha}_n,
\qquad \bm{\sigma}=\sum_{n=1}^{N-1} \sigma_n \bm{\mu}_n, 
\ee
so that $\phi_n$, $\sigma_n$ are $2\pi$-periodic scalars, where $\bm{\alpha}_{n}$ denote the positive simple roots, and $\bm{\mu}_{n}$ denote the fundamental weights. 
In addition, the remnant of non-Abelian gauge invariance exists, and we have to perform the gauge identification by the Weyl group, $W_{\mathrm{SU}(N)}=S_N$, which gives the permutations of the vector components. 
Therefore, the $(\bphi,\bsigma)$ takes values in the following space~\cite{Argyres:2012ka},
\begin{equation}
  \frac{\mathbb{R}^{N-1}/2\pi\Lambda_r
    \times\mathbb{R}^{N-1}/2\pi\Lambda_w}{S_N}\,.
\end{equation}
Here $\Lambda_r=\sum_n 2\pi \mathbb{Z} \bm{\alpha}_n$ and $\Lambda_w=\sum_n 2\pi\mathbb{Z} \bm{\mu}_n$ are the root and weight lattices, respectively. 

It turns out to be useful to fix the gauge for the Weyl permutation, $S_N$, using $\bm{\phi}$. 
Indeed, we can characterize the physically inequivalent classical vacua of $\bm{\phi}$ by the conditions; 
\be
\bm{\alpha}_n\cdot \bm{\phi}>0\;\; (n=1,\ldots, N-1), \quad -\bm{\alpha}_0\cdot \bm{\phi}<2\pi,
\label{eq:fixing}
\ee
which is called the fundamental Weyl chamber. Here, $\bm{\alpha}_0=-(\bm{\alpha}_1+\cdots+\bm{\alpha}_{N-1})$ is the Affine simple root. 
In components, these conditions can be written as 
\bea
&&\phi_{n+1}+\phi_{n-1}<2\phi_n \quad (n=1,\ldots,N-1),\nonumber\\
&&\phi_1+\phi_{N-1}<2\pi,
\eea
in which we regard $\phi_0=\phi_N=0$ and $0\le \phi_n < 2\pi$. 
The Weyl chamber for all the simply-connected and simple gauge groups can be found in Appendix~B of Ref.~\cite{Argyres:2012ka}.

Next, let us explain how the $0$-form symmetries discussed in Sec.~\ref{sec:symmetries} act on these fields $(\bm{\phi},\bm{\sigma})$. 
The center and $\mathcal{CP}$ symmetry transformations are generated by
\begin{equation}\label{eq:symmetry1}
  \begin{array}{ll}
  \text{Center (``naive''):} \quad \left(\bphi,\bsigma\right)\to\left(\bphi+2\pi\bm{\mu}_1,\bsigma\right),\\
  \CP: \qquad
  \begin{cases}
    \theta=0:\quad \left(\bphi,\bsigma\right)\to\left(\bphi, - \bsigma\right),\\
    \theta=\pi:\quad \left(\bphi,\bsigma\right)\to\left(\bphi, - \bsigma - \bphi\right).
  \end{cases}
  \end{array}
\end{equation}
Here, we note that the above definition of the center transformation is naive, as it does not respect the gauge fixing condition (\ref{eq:fixing}). 
In order to make the center transformation being closed inside the Weyl chamber, we combine it with the cyclic Weyl permutation, $\perm\in S_N$~\cite{Argyres:2012ka}: 
\begin{equation}\label{eq:symmetry2}
    \begin{array}{ll}
    &\text{Center:} \qquad\qquad\;\, \left(\bphi,\bsigma\right)\to\left(\perm\bphi+2\pi\bm{\mu}_1,\perm\bsigma\right),\\
    &\CP: \quad
    \begin{cases}
         \theta=0:\quad \left(\bphi,\bsigma\right)\to\left(\bphi,-\bsigma\right),\\
         \theta=\pi:\quad \left(\bphi,\bsigma\right)\to\left(\bphi,-\bsigma-\bphi\right).
    \end{cases}
    \end{array}
\end{equation}
The action of $\perm$ is given by $\perm\bm{\alpha}_n=\bm{\alpha}_{n+1}$, where the label is understood in mod $N$, and $\perm\bm{\mu}_n=\bm{\mu}_n-(\bm{\alpha}_1+\cdots+\bm{\alpha}_n)$. 
Using the Weyl vector, $\bm{\rho}=\sum_{n}\bm{\mu}_n$, the center-symmetric configuration for $\bm{\phi}$ is given by 
\be
\phic={2\pi\over N}\bm{\rho},
\label{eq:phisup}
\ee
as we can check that $\perm\bm{\rho}=\bm{\rho}-N\bm{\mu}_1$. 
Indeed, Eq.~\eqref{eq:phisup} is realized as the vacuum configuration for the SUSY case with $m=0$. 

Let us now consider the nonperturbative effect for the bosonic effective potential. 
It is convenient to introduce complex scalar fields;
\begin{equation}\label{eq:z}
  \bz \equiv \im \biggl[ \left({\theta\over 2\pi}+{\im\over \alpha_s}\right)(\bm{\phi}-\phic)+\bm{\sigma}\biggr] \,,
\end{equation}
then the kinetic term (\ref{eq:kinetic}) can be compactly expressed as ${\alpha_s\over 4\pi L}|\diff \bm{z}|^2$. 
The $4$D YM instanton splits into $N$ monopole-instantons in this setup~\cite{Lee:1997vp, Lee:1998bb, Kraan:1998kp,Kraan:1998pm,Kraan:1998sn}, 
and those monopole-instantons consist of $(N-1)$ BPS monopoles with the magnetic charge, $\bm{\alpha}_n$ ($n=1,\ldots,N-1$), and a KK monopole with the magnetic charge, $\bm{\alpha}_0$. 
The existence of KK monopole-instanton is crucial for the mass-gap generation~\cite{Davies:2000nw, Unsal:2007jx,Unsal:2007vu,Unsal:2008ch} by the Polyakov-type mechanism~\cite{Polyakov:1976fu}, which does not occur for the genuine $3$D SYM theory~\cite{Affleck:1982as}. 
Let us introduce the monopole-instanton operators as 
\be
M_i(x) = \exp\biggl[\bm{\alpha}_i\cdot\bm{z}(x)+\im\,{\theta\over N}\biggr]\,. 
\label{eq:monopole}
\ee
At the SUSY point, $m=0$, the monopole-instantons cannot contribute to the effective potential by themselves as they carry two fermionic zero modes by the index theorem~\cite{Poppitz:2008hr}. 
Therefore, the BPS and KK monopole-instanton vertices are given by
\be
\exp\left(-{2\pi\over N\alpha_s}\right) M_i(x)\, [\bm{\alpha}_i\cdot \bm{\lambda}(x)]^2, 
\ee
where $\bm{\lambda}$ denotes the diagonal components of gluinos. 

Let us now write down the bosonic effective potential for the softly-broken $\calN=1$ SYM theory on small $\mathbb{R}^3\times S^1$. 
As we noted, the monopoles $M_i$ cannot contribute at the SUSY point, $m=0$, and thus the leading contribution comes out of ``bions''~\cite{Unsal:2007jx,Unsal:2007vu,Unsal:2008ch}, which are composites of $M_i$. 
They should not carry the topological charge, so the candidates are $M_i^*M_j$. When $i=j$, they do not have the magnetic charge either, so they are called neutral bions. When $i\not=j$, they have the magnetic charge $\bm{\alpha}_j-\bm{\alpha}_i$, so they are called magnetic bions. 
Away from the SUSY point, the monopoles $M_i$ can also contribute to the bosonic potential of the order of $m$. 
These contributions make up the following effective potential~\cite{Davies:2000nw, Unsal:2007jx,Unsal:2007vu,Unsal:2008ch,Anber:2014lba}, 
\bea
  \frac{\mathcal{V}}{V_0} \,
  &&= \sum_{i,j=0}^{N-1}(\balpha_i\cdot\balpha_j)  M_i^* M_j \nonumber\\
  &&\quad - {\gamma\over 2} \sum_{i=0}^{N-1} \biggl[ 1 - \frac{N \alpha_s}{4\pi} \ln(M_i^*M_i) \biggr] (M_i+M_i^*),
  \label{eq:potential}
\eea
where the first term comes out of bions, and the second term is the monopole contribution. 
It is interesting to notice that, since $\bm{\alpha}_i\cdot \bm{\alpha}_j=2\delta_{i,j}-\delta_{i,j+1}-\delta_{i,j-1}$, the magnetic bions $M^*_i M_j$ contribute if and only if $j=i\pm 1$. 
We have normalized the potential using the bion amplitude,
\begin{equation}
  V_0 \equiv \frac{9 N^2}{64\pi^3 }
  \frac{L^3 \Lambda^6}{\alpha_s} \,,
\end{equation}
which characterizes the superpotential scale at $m=0$.  As we
discussed around Eq.~\eqref{eq:def_gamma_rough}, $\gamma$ in
Eq.~\eqref{eq:potential} is a proxy of the temperature $T$ or the
inverse system size $L^{-1}$ as a control parameter to probe the phase
structure, which is defined as\footnote{Here, we note that our definition of $V_0$ is a half of $V^0_{\mathrm{bion}}$ used in Ref.~\cite{Poppitz:2012nz}. As a result, the dimensionless temperature $\gamma$ is related to their parameter, $c_m$, in Ref.~\cite{Poppitz:2012nz} by $\gamma=2c_m$. }
\begin{equation}\label{eq:gamma}
  \gamma \equiv \frac{32\pi^2 }{3 N^2}
  \frac{m}{L^2 \Lambda^3} \,. 
\end{equation}

In addition to the semiclassical potential~\eqref{eq:potential} there is
another contribution from perturbative loops, i.e., a GPY-like
potential~\cite{Gross:1980br}.  The perturbative
contribution is, however, suppressed by SUSY, which can be seen from
\begin{equation}
  \frac{\VGPY}{V_0} = -\gamma^2 \frac{\alpha_s N^2}{4\pi}
  \sum_{\balpha\in\Phi^+} B_2 \biggl( \frac{\balpha\cdot\bphi}{2\pi}
  \biggr)\,,
\end{equation}
where the $\balpha$ summation runs over positive roots $\Phi^+$, and $B_2(x)=x^2-x+1/6$ is the second Bernoulli polynomial. 
We are interested in the $\mathcal{O}(1)$ fluctuation of $M_i$ to analyze the effective potential~\eqref{eq:potential}, and this means that the $\bm{\phi}$ fluctuation should be $\bm{\phi}-\phic\sim \mathcal{O}(\alpha_s)$ according to Eq.~\eqref{eq:z}. 
As a consequence, we find,
\be
{\VGPY(\bm{\phi})-\VGPY(\phic)\over V_0}\sim \mathcal{O}(\alpha_s^2 N^2). 
\ee
For $\mathrm{SU}(N)$, we can confirm that the perturbative potential is actually more suppressed as $\mathcal{O}(\alpha_s^3 N^2)$~\cite{Poppitz:2012sw,Poppitz:2012nz}. 
Therefore, we neglect the perturbative contribution in the following, and use Eq.~\eqref{eq:potential} for the computations.

\section{Simultaneous Deconfinement and $\mathcal{CP}$ Restoration for $G=\mathrm{SU}(N\ge 3)$}\label{sec:other}

In this section, we discuss the general phase structure of the effective theory we introduced in Sec.~\ref{sec:potential}. 
Although we mainly work on the $\mathrm{SU}(N)$ gauge groups for $N\ge 3$ in this section, the generalization to other gauge groups is straightforward~\cite{Anber:2014lba}, which will be discussed in Appendix~\ref{app:nonSU}. 
Especially, we shall see that the simultaneous deconfinement and $\mathcal{CP}$ restoration occurs for all the gauge groups except $\mathrm{SU}(2)$. 
We will discuss the $\mathrm{SU}(2)$ case in the next section.

Instead of keeping the original variables, $(\bphi,\, \bsigma)$, for computation, it turns out to be useful to work with $N$-complex scalars $M_i$, which are the monopole-instanton operators introduced in Eq.~\eqref{eq:monopole}. 
These $N$ fields, $M_i$, are not all independent of each other but there is a constraint:
\be
\label{eq:constraint_m}
\prod_{i=0}^{N-1}M_i = \exp(\im \theta). 
\ee
Under the constraint \eqref{eq:constraint_m}, the $N$-complex fields $M_i$ have the one-to-one correspondence to $(\bphi,\, \bsigma)$.

Since we perform our computation with $M_i$, center and $\mathcal{CP}$ symmetries (\ref{eq:symmetry2}) should also be translated in the language of $M_i$. 
The result is surprisingly simple: 
\be
\label{eq:symmetry3}
    \begin{array}{ll}
    &\text{Center:} \quad M_i\to M_{i+1},\\
    &\CP: \qquad  M_i\to M_i^*. 
    \end{array}
\ee
In terms of the monopole variables, the appearance of $\mathcal{CP}$ only at $\theta=0$ or $\theta=\pi$ can be understood as follows: Only when $\theta=0$ or $\theta=\pi$, the constraint (\ref{eq:constraint_m}) is real, and then the $\mathcal{CP}$ transformation, $M_i\to M_i^*$, is consistent with the constraint. 

Let us point out that this theory correctly reproduces the 't~Hooft anomaly of YM theory~\cite{Tanizaki:2019rbk}. 
Even without any knowledge about the effective potential, the target space of these infrared degrees of freedom, $M_i\Leftrightarrow (\bphi,\, \bsigma)$, already knows about the 't~Hooft anomaly at $\theta=\pi$ to some extent.
To see it, let us try to find out the center and $\mathcal{CP}$ symmetric points. The requirement of center symmetry is that 
\be
M_0=M_1=\ldots =M_{N-1} \equiv M. 
\ee
Therefore, the constraint (\ref{eq:constraint_m}) is solved under the center-symmetric condition as $M_i=M$ for all $i=0,1,\ldots,N-1$ with
\be
M=\exp\left(\im\,{\theta+2\pi k\over N}\right).
\label{eq:confined_vacua}
\ee
Here, $k=0,1,\ldots, N-1$, so there are $N$ isolated center-symmetric points. This is a manifestation of $N$-branch structure of confining vacua, expected in the large-$N$ counting~\cite{Witten:1980sp, Witten:1998uka}. 
We can readily check that none of them is $\mathcal{CP}$ symmetric when $N$ is even, so there cannot exist a unique gapped vacuum in the weakly-coupled regime. 
When $N$ is odd, there is a center and $\mathcal{CP}$ symmetric point for each of $\theta=0$ and $\theta=\pi$, but their labels $k$ do not coincide for different $\theta$'s, which is a manifestation of global inconsistency. 

When $m=0$, all of the $N$ configurations in Eq.~\eqref{eq:confined_vacua} are realized as actual vacua. This $N$-degeneracy for the SUSY point is indeed understood as a consequence of spontaneously broken discrete chiral symmetry, often called $R$-symmetry, and they are characterized by the gluino condensate, 
\be
\bigl\langle \tr(\lambda^2)\bigr\rangle = 3N \Lambda^3 \exp\left(\im\,{\theta+2\pi k\over N}\right),
\ee 
with $k=0,1,\ldots, N-1$. 
The existence of $N$ vacua is ordered by a similar mixed 't Hooft anomaly as Eq.~\eqref{eq:anomaly}. The symmetry transformation is generated by $M_i\to M_i \exp\left(2\pi \im /N\right)$. When $m\neq 0$, the second term in the potential \eqref{eq:potential} destroys $R$-symmetry and lifts the degeneracy. For $0\leq\theta<\pi$ the $k=0$ branch survives while for $\pi<\theta\leq 2\pi$ the $k=N-1$ branch survives. At $\theta=\pi$, these two branches are degenerate, which gives the anticipated first-order phase transition of $\CP$-breaking.

We have discussed properties for the center-symmetric, or confining, vacua, and let us next describe the deconfined phase. 
To get an insight, we set $\bm{\phi}=\bm{\sigma}=0$, which is a typical center-broken configuration in the perturbative analysis, even though we lose the validity of Abelinized description at that point. 
The monopole-instanton operators (\ref{eq:monopole}) become
\be
\label{eq:deconfined}
M_0=\exp\biggl[-{2\pi\over N\alpha_s}(N-1)+\im \theta\biggr] , \quad M_{i\not=0}=\exp\left({2\pi\over N\alpha_s}\right). 
\ee
In this case we see $|M_0| \ll |M_{i\not=0}|$. As we set $\theta=\pi$, we find $M_0<0$ while $M_i>0$ for all $i\neq0$. We can guess that these qualitative behaviors of $M_i$ are also true for the actual deconfined vacua, at least for deeply deconfined vacua. That is, in such a deconfined vacuum, one monopole is suppressed compared with others, and at $\theta=\pi$, the suppressed one has a negative amplitude, while others take positive values. Since all the amplitudes are real, $\CP$ symmetry would be manifested.

Now, let us think about the phase transition between the confined and deconfined phases. 
We note that the deconfinement transition is of the first order for $\mathrm{SU}(N)$ gauge groups with $N\geq3$~\cite{Poppitz:2012nz}. 
Hence, the simplest scenario for the deconfinement transition is just to exchange the confined vacua by the deconfined vacua with the qualitative features we have guessed above. If this is the case, then the deconfined vacua are $\mathcal{CP}$ symmetric at $\theta=\pi$, so we should get,
\be
\gammadec(\theta=\pi)=\gammaCP.
\ee
That is, the deconfinement and $\mathcal{CP}$ restoration occur simultaneously at $\theta=\pi$. 

Although this is just a naive guess, we can confirm expected features by solving Eq.~\eqref{eq:potential}. 
Since all the equations we need to solve are too intricate to treat analytically,
we evaluate them numerically setting $\alpha_s=0$ in the potential \eqref{eq:potential}. Neglecting $\calO(\alpha_s)$ terms does not affect the qualitative behaviors due to the first-order nature of the deconfinement transition. 
We list the numerical results of the deconfinement phase transition for $\mathrm{SU}(N\ge 3)$ up to $N=10$ in Table~\ref{tab:A}.  The table exhibits $\gammadec(\pi)$ as well as the expectation values of $M_i$ in one of the deconfined vacua at $\gammadec(\pi)$. Other deconfined vacua can be obtained by the broken center transformation \eqref{eq:symmetry3}. 

\begin{table}
  \centering
\begin{tabular}{|c|c|l|c|}
     \hline $G$ & $\mathcal{Z}(G)$ & \multicolumn{1}{c|}{$M_{i}$} & $\gammadec(\pi)$ \\\hline\hline 
     $\mathrm{SU}(3)$ & $\mathbb{Z}_3$ & \begin{tabular}{l}
          -0.167, 2.451, 2.451
     \end{tabular} & 4.235 \\\hline 
      $\mathrm{SU}(4)$ & $\mathbb{Z}_4$ & \begin{tabular}{l}
          -0.077, 2.130, 2.849, 2.130
     \end{tabular} & 2.564 \\\hline 
     $\mathrm{SU}(5)$ & $\mathbb{Z}_5$ & \begin{tabular}{l}
          -0.039, 1.835, 2.756, 2.756,\\ 1.835
     \end{tabular} & 1.711 \\\hline 
     $\mathrm{SU}(6)$ & $\mathbb{Z}_6$ & \begin{tabular}{l}
          -0.021, 1.600, 2.560, 2.879,\\ 2.560, 1.600
     \end{tabular} & 1.220 \\\hline 
     $\mathrm{SU}(7)$ & $\mathbb{Z}_7$ & \begin{tabular}{l}
          -0.011, 1.414, 2.357, 2.827,\\ 2.827, 2.357, 1.414
     \end{tabular} & 0.914 \\\hline 
     $\mathrm{SU}(8)$ & $\mathbb{Z}_8$ & \begin{tabular}{l}
          -0.006, 1.267, 2.171, 2.713,\\ 2.893, 2.713, 2.171, 1.267
     \end{tabular} & 0.709 \\\hline 
     $\mathrm{SU}(9)$ & $\mathbb{Z}_9$ & \begin{tabular}{l}
          -0.003, 1.147, 2.007, 2.580,\\ 2.866, 2.866, 2.580, 2.007,\\ 1.147
     \end{tabular} & 0.566 \\\hline 
     $\mathrm{SU}(10)$ & $\mathbb{Z}_{10}$ & \begin{tabular}{l}
          -0.002, 1.049, 1.864, 2.446,\\ 2.795, 2.911, 2.795, 2.446,\\ 1.864, 1.049
     \end{tabular} & 0.462 \\\hline 
\end{tabular}
  \caption{Numerically evaluated $\gammadec(\pi)$ for the gauge groups $\mathrm{SU}(N)$ with $N\geq 3$. 
The expectation values of $M_i$ are given in one of the deconfined vacua at $\gamma=\gammadec(\pi)$. 
  }
  \label{tab:A}
\end{table}

We performed the numerical calculations with accuracy to the 9th decimal place. 
We exhibit in the table at most 4 significant digits for $\Re (M_i)$, while $\Im (M_i)$ is indeed identically zero up to $10^{-10}$. The features from our naive guess turn out to be correct even near the phase transition point: One monopole is suppressed and has a negative amplitude, while others are positive and unsuppressed.
Another feature is that many numbers appear in pairs, which suggests an unbroken $\mathbb{Z}_2$ symmetry. It is indeed the charge conjugation symmetry, $\mathcal{C}: M_i\mapsto M_{N-i}$.

Recalling $\CP$ symmetry is given by complex conjugation of $M_i$, we infer that the deconfined vacua at $\gammadec(\pi)$ are $\CP$ symmetric, from the vanishing imaginary parts of $M_i$ in Table~\ref{tab:A}. That is to say, $\CP$ symmetry is restored simultaneously when center symmetry starts to be broken, i.e., $\gammadec(\pi)=\gammaCP$ is observed for $\mathrm{SU}(N)$ with $N\ge 3$. 

In Appendix~\ref{app:nonSU}, we also study the deconfinement transition at $\theta=\pi$ for the non-$\mathrm{SU}(N)$ gauge groups, $G$. 
The deconfinement transition turns out to be always of the first order when $G\not=\mathrm{SU}(2)$~\cite{Poppitz:2012nz, Anber:2014lba}, and we confirm that $\gammadec(\pi)=\gammaCP$ also for the non-$\mathrm{SU}(N)$ gauge groups up to the rank-$9$ groups, including $\mathrm{Spin}(N\le 19)$, $\mathrm{Sp}(N\le 9)$, $E_{6,7,8}$, $F_4$, and $G_2$. 
In Appendix~\ref{app:largeN}, we solve the deconfined phase of (\ref{eq:potential}) analytically in the large-$N$ limit of $\mathrm{SU}(N)$ groups, and we again confirm the correctness of above discussions. 
We thus conclude that deconfinement and $\mathcal{CP}$ restoration occur simultaneously at $\theta=\pi$ for all the gauge groups $G\not=\mathrm{SU}(2)$. 
We also anticipate optimistically that this equality might be extrapolated to the pure thermal Yang-Mills theory, namely $\Tdec(\pi)=\Tcp$ for all $G\not=\mathrm{SU}(2)$, which awaits further numerical studies.

\section{The deconfined and $\CP$-broken phase in the SU(2) case}\label{sec:SU(2)}

In this section, we will evaluate the phase diagram on the $\theta$-$\gamma$ plane in the case of $G=\mathrm{SU}(2)$. Now the deconfinement transition is of second order~\cite{Poppitz:2012sw}. According to the discussion around Eq.~\eqref{eq:confined_vacua}, no point is simultaneously center and $\CP$-symmetric. 
Since we are studying the weak-coupling regime, this immediately implies the existence of a finite window $\gammadec(\pi) < \gammaCP$ as well as a novel phase of deconfinement and $\CP$-breaking. It is so because a second-order transition requires a continuous motion of the vacua. As a result, we must have a phase with neither center nor $\CP$ symmetry between two vacua with one and the other symmetries.
We will justify this argument by a quantitative evaluation of $\gammadec(\theta)$ and $\gammaCP$.

From now on we will work with the original variables, $(\bphi,\,\bsigma)$.
The expectation value of $\bphi$ in the center-symmetric vacua now reads $\phic = \pi\bmu_1$. We then express $(\bphi,\,\bsigma)$ in components:
\begin{equation}
    \bphi \equiv \phic + \alpha_s\varphi \bmu_1 \,,\quad
    \bsigma \equiv \sigma \bmu_1 \,.
\end{equation}
The components $\varphi$ and $\sigma$ are real variables and have the following relation to $\bz$:
\begin{equation}
    \varphi = -\balpha_1\cdot \Re \bz \,,\quad\sigma = \balpha_1\cdot \Im \bz - \frac{\theta\alpha_s}{2\pi}\varphi\,.
\end{equation}
Their target spaces can be regarded as $\varphi\in\mathbb{R}$ and $\sigma\in\mathbb{R}/2\pi \mathbb{Z}$.

The symmetry transformations \eqref{eq:symmetry2} now read:
\begin{equation}
    \begin{array}{ll}
    \text{Center Trans.:} \quad (\varphi,\sigma) \to (-\varphi, -\sigma) \,,\\
    \text{$\CP$ Trans.:} \\\quad \begin{cases}
        \theta=0:\quad(\varphi,\sigma) \to (\varphi,-\sigma)\,,\\
        \theta=\pi:\quad(\varphi,\sigma) \to (\varphi,-\sigma-\pi-\alpha_s\varphi)\,.
    \end{cases}
    \end{array}
\end{equation}
These symmetries are both $\mathbb{Z}_2$ here. It is clearly seen that no $(\varphi,\sigma)$ stays invariant under two $\mathbb{Z}_2$'s simultaneously at $\theta=\pi$, which is consistent with the mixed 't~Hooft anomaly. The SU(2) potential \eqref{eq:potential} as a function of $\varphi$ and $\sigma$ for a given $\theta$ reads:
\begin{equation}
  \begin{split}
  &\frac{\Vbos(\varphi,\sigma;\theta)}{V_0}
   = 4\cosh(2\varphi) - 4\cos\left(2\sigma + \frac{\theta\alpha_s}{\pi}\varphi\right) \\
  &\qquad-\gamma \biggl[ \biggl( 1+\frac{\alpha_s}{\pi}\varphi\biggr)
  e^{-\varphi}\cos\left(\sigma+\frac{\theta}{2}+\frac{\theta\alpha_s}{2\pi}\varphi\right)\\
  &\qquad\quad + \biggl( 1-\frac{\alpha_s}{\pi}\varphi
  \biggr) e^\varphi \cos\left(\sigma-\frac{\theta}{2}+\frac{\theta\alpha_s}{2\pi}\varphi\right) \biggr]\,.
\end{split}
\label{eq:su2poten}
\end{equation}
Center and $\CP$ symmetries are evident from the above potential.

We determine $\varphi$ and $\sigma$ from a condition to minimize Eq.~\eqref{eq:su2poten}. As long as $\gamma$ is sufficiently small (we will soon evaluate the threshold) the system stays in the confined phase and center symmetry is unbroken. Then, according to Eq.~\eqref{eq:confined_vacua}, $\varphi=0$ and $\sigma=0,\pi$ should be energetically favored, which simplifies the analysis.
The energy in the confined phase with $\varphi=0$ is then,
\begin{equation}
    \frac{\Vbos(\varphi=0,\sigma;\theta)}{V_0} 
   =
     8- 8\cos^2 \sigma - 2 \gamma \cos \left({\theta\over 2}\right) \cos \sigma. 
   \end{equation}
From this expression it is clear that $\cos\sigma=+1$ lowers the energy as long as $\cos(\theta/2)> 0$, while $\cos\sigma=-1$ is energetically favored for $\cos(\theta/2)< 0$. That is, we find an anticipated two-branch structure:
\begin{equation}
  \sigma(\theta) = \begin{cases}
    \displaystyle 0 & (0\le\theta < \pi)\,, \\[0.3em]
    \displaystyle \pi & (\pi < \theta \leq 2\pi)\,.
  \end{cases}
  \label{eq:sigma_conf}
\end{equation}
This certainly gives the branch structure we discussed in Eq.~\eqref{eq:confined_vacua}, and indicates a first-order phase transition with spontaneous $\mathcal{CP}$ breaking at $\theta=\pi$ in the confined phase.  We note that the nature of the phase transition is insensitive to $\gamma$ if we work in the confined phase. 

\begin{figure}
  \includegraphics[width=0.9\columnwidth]{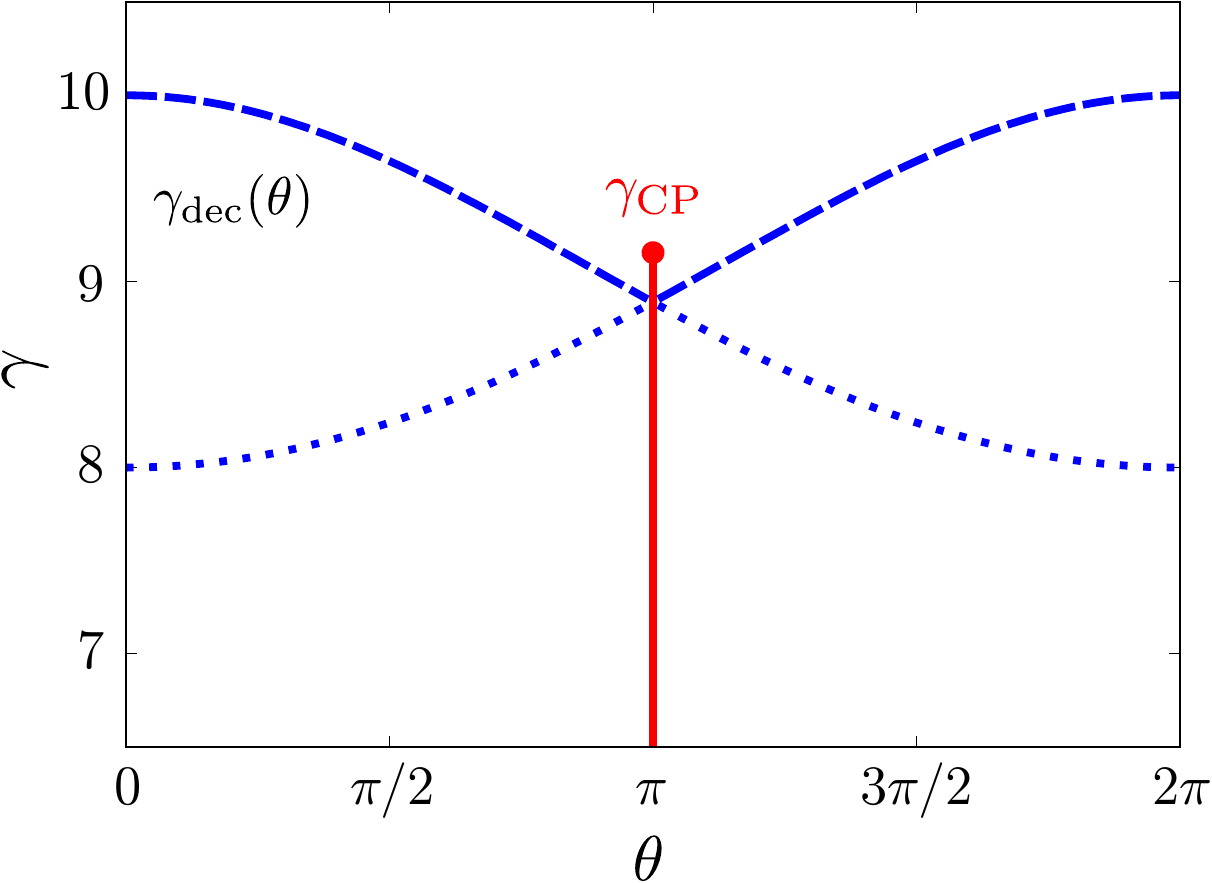}
  \caption{SU(2) phase structure on the $\theta$-$\gamma$ plane for $\alpha_s/\pi=0.1$. The dashed lines represent the deconfinement boundary given by $\gamma_{\mathrm{dec}}(\theta)$ in Eq.~\eqref{eq:gammadec} with the metastable branch drawn by the dotted lines. The first-order phase boundary associated with the CP breaking at $\theta=\pi$ is shown by the solid (red) line which terminates at $(\theta,\gamma)=(\pi,\gamma_{\mathrm{CP}})$.}
  \label{fig:phase}
\end{figure}

Let us now consider a critical $\gammadec$, above which center symmetry is broken.  For the SU(2) case the deconfinement phase transition is of second order, so that the potential at $\gamma=\gammadec$ must become flat around $\varphi=0$. If we approach the phase boundary from $\gamma < \gammadec$, we can keep using $\sigma(\theta)$ in Eq.~\eqref{eq:sigma_conf}.  We can impose the flatness condition on the potential in a form that the Hessian of the potential should vanish, i.e., $(\partial^2\Vbos/\partial\varphi^2)(\partial^2\Vbos/\partial\sigma^2)-(\partial^2\Vbos/\partial\varphi\partial\sigma)^2=0$.
This leads to the following solution,
\begin{equation}
  \gamma_{\text{dec}}(\theta)
  = \begin{cases}
    \displaystyle \frac{8}{\displaystyle 1 - \frac{\alpha_s}{\pi}
      \bigl[ 1+\cos(\theta/2) \bigr]} \quad (0\le\theta<\pi)\,,
    \\[1em]
    \displaystyle \frac{8}{\displaystyle 1 - \frac{\alpha_s}{\pi}
      \bigl[ 1-\cos(\theta/2) \bigr]} \quad (\pi<\theta<2\pi)\,,
  \end{cases}
  \label{eq:gammadec}
\end{equation}
which draws the phase boundary of deconfinement (see dashed lines in
Fig.~\ref{fig:phase}).

We notice that $\gammadec(\theta)$ decreases as $\theta$ approaches $\pi$ from $0$. This phenomenon can be qualitatively understood as follows: The bions tend to confine the theory, while the monopole-instantons tend to deconfine the theory. SUSY helps us to induce bion contributions, so the chiral anomaly by the gluino destroys the influence of $\theta$. However, $\theta$ gives a phase to the KK monopole-instanton amplitude and would bring significant constructive interference among monopole-instantons. Therefore, $\theta$ would increase the deconfining force in the system, which results in the decreasing behavior of $\gammadec(\theta)$. 

The results up to now in this section are consistent with Ref.~\cite{Poppitz:2012nz}.
To complete the phase diagram the last piece we need is the determination of $\gammaCP$, where the $\CP$-breaking first-order phase transition is terminated. This needs to be done together with finding the vacua there. Since $\gammaCP$ is located at an endpoint of a first-order phase boundary, this is a critical point of second order. Therefore, we can find this phase transition similarly by requiring a vanishing gradient and a vanishing Hessian. Now $\CP$ can simplify our tasks. Firstly, because $\CP$ is restored at $\gammaCP$, we directly know $\sigma=\frac{\pi}{2}-\frac{\alpha_s}{2}\varphi$ or $\frac{3\pi}{2}-\frac{\alpha_s}{2}\varphi\mod 2\pi$ there. Secondly, among the vanishing gradient equations, i.e., $\partial\Vbos/\partial\varphi = 0$ and $\partial\Vbos/\partial\sigma = 0$, the latter is automatically satisfied due to the $\CP$ symmetry. Lastly, $\CP$ symmetry also tells us that the flat direction must be along $\sigma$, which reduces a vanishing Hessian condition to just $\partial^2 \Vbos/\partial\sigma^2 = 0$.

We start from solving $\partial\Vbos/\partial\varphi = 0$ at $\theta=\pi$ and $\sigma=\frac{\pi}{2}-\frac{\alpha_s}{2}\varphi$, which results in a function $\varphi(\gamma)$. We cannot find an analytical expression of $\varphi(\gamma)$, so we have expanded it
with respect to $\alpha_s/\pi \ll 1$. The solution up to
$O(\alpha_s/\pi)$ is
\begin{equation}
  \begin{split}
    \varphi(\gamma) &\simeq \arcsinh(\gamma/8) \\
    &\quad - \frac{\alpha_s/\pi}{\sqrt{1+64/\gamma^2}}
    \biggl[ 1+\frac{\arcsinh(\gamma/8)}{\sqrt{1+64/\gamma^2}} \biggr]\,.
  \end{split}
\end{equation}
With this solution we can approximately solve $\partial^2 \Vbos/\partial\sigma^2 = 0$, up to
$O(\alpha_s/\pi)$ to find,
\bea
  \gammaCP &\simeq& 8 \biggl[ 1 + \frac{1}{2}
  \biggl( 1 + \frac{3}{\sqrt{2}}\arcsinh (1) \biggr)
  \frac{\alpha_s}{\pi} \biggr]\nonumber\\
  &\simeq& 8 \biggl( 1 + 1.43\frac{\alpha_s}{\pi} \biggr)\,,
\eea
which is surely larger than the deconfinement critical value,
\begin{equation}
  \gammadec(\theta=\pi) \simeq 8\left(1+\frac{\alpha_s}{\pi}\right)\,.
\end{equation}
Hence, as we promised, a finite window $\gammadec(\pi)<\gamma<\gammaCP$ appears, with the window size of $\calO(\alpha_s)$
(see the brown solid line in Fig.~\ref{fig:phase}).

We have now completed the phase diagram for $G=\mathrm{SU}(2)$ and found that there is a novel phase of deconfinement with $\CP$ breaking in a finite window $\gammadec(\pi)<\gamma<\gammaCP$. This novel phase has four degenerate vacua, while any other place in the phase diagram has at most two degenerate vacua. These vacua will be carefully considered in the next section. 

From Fig.~\ref{fig:phase}, we see that the phase structure has an unusual feature, namely, the first-order transition line intersects with the second-order transition line. This peculiar behavior is rarely seen in literature, as far as we know. From now on, it should be incorporated as a recognized member in a zoo of various phases. 

We could optimistically expect that these interesting behaviors may be extrapolated to the thermal pure $\mathrm{SU}(2)$-gauge theory, with a replacement of $\gamma$ by $T$. However, one should be careful of the peculiarity of $\mathrm{SU}(2)$ and we would like to reserve other possibilities. For example, both the deconfinement and $\CP$-restoration phase transitions might terminate at zero temperature in pure $\mathrm{SU}(2)$ gauge theory, and the theory might become massless around $\theta=\pi$~\cite{Gaiotto:2017yup}. 
Another interesting scenario would be the 't~Hooft's oblique confinement scenario~\cite{tHooft:1981bkw, Cardy:1981qy}, which concludes the deconfinement for the $\mathrm{SU}(2)$ YM theory at $\theta=\pi$.  
We cannot make any decisive statement about a favored scenario for pure YM theory just from the symmetry constraint and from our present analysis in the weak-coupling regime. 
Nevertheless, because the pure YM theory is still far from unraveled, it would be an important step to make exhaustive studies of possible scenarios and our present calculations illuminate a novel scenario that has not been known before.

\section{1-form center symmetry and domain walls in the novel phase}\label{sec:junction}

In the novel deconfined and $\mathcal{CP}$-broken phase, there are four discrete vacua related by the broken center and broken $\mathcal{CP}$ symmetries. 
Therefore, we can consider more variations of domain wall excitations and domain-wall junctions than in other phases.
In this section, we consider the 1-form center symmetry on the domain wall and the junction in this novel phase. 

We should start with reviewing the spatial Wilson loop in the effective description. 
In the infrared effective theory, we have a $1$-form $\mathrm{U}(1)^{N-1}$ topological symmetry ordered by $\bsigma\sim\bsigma+2\pi\Lambda_w$ (recall that $\bphi$ cannot deviate far from the center-symmetric point $\phic$). An order operator is a loop defect, around which $\bsigma$ has a $2\pi\bmu$ monodromy with a $\bmu\in\Lambda_w$. 
The $1$-form $\mathrm{U}(1)^{N-1}$ symmetry acts as the phase rotations of such a line defect. 
Recall that $\bsigma$ is the Abelianized dual $3$D gauge field, so the loop defect is nothing but a spatial Wilson loop with an electric charge $\bmu$, and the subgroup $\mathbb{Z}_N\subset \mathrm{U}(1)^{N-1}$ of this 1-form symmetry matches the 1-form center symmetry of $\mathrm{SU}(N)$ YM theory. The infrared enhancement of $1$-form symmetry from $\mathbb{Z}_N$ to $\mathrm{U}(1)^{N-1}$ indicates the Abelianization.

We can find the area law of spatial Wilson loops by seeing that they create the confining-string excitations.
A Wilson loop imposes the boundary condition of $\bsigma$ so that it should have a nontrivial winding around the loop. 
Solving the infrared equation of motion under that boundary condition, we can explicitly see the configuration of the wall bounded by a Wilson loop~\cite{Anber:2015kea, Polyakov:1976fu}, which is nothing but a sheet of confining strings. 
As a result, the Wilson loop shows the area law, and its string tension is proportional to the wall tension, which also justifies the integrity of the 1-form center symmetry.

\begin{figure}[t]
  \includegraphics[width = 0.72 \columnwidth]{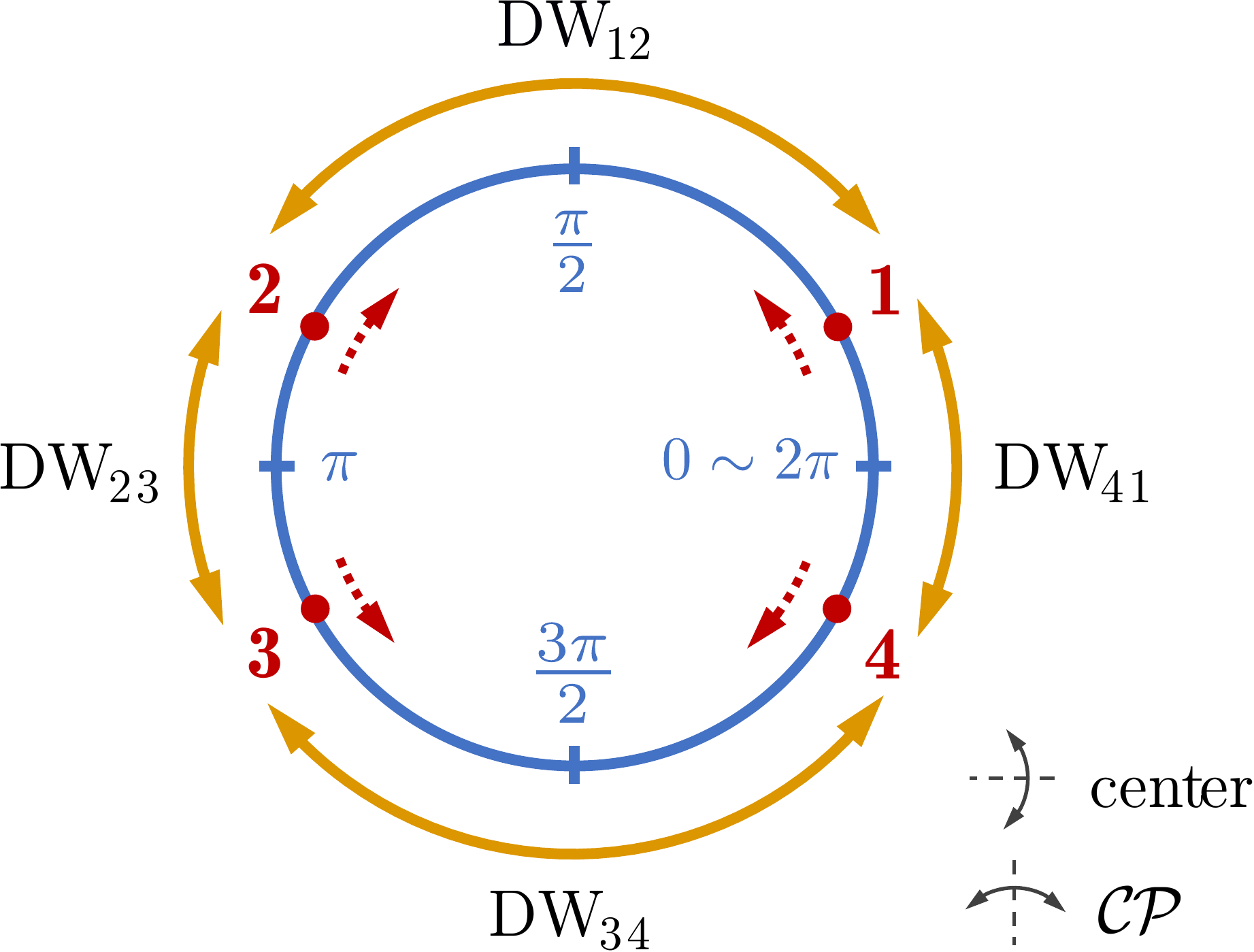}
  \caption{The vacua and domain walls in the novel phase. The blue circle represents the value of $\sigma\sim \sigma+2\pi $. The red dots denote the vacua from 1 to 4, and the red dashed arrows indicate the moving direction of these vacua as $\gamma$ goes larger. Between the neighboring vacua, the yellow two-sided arrows indicate the domain walls. The (0-form) center symmetry corresponds to the horizontal reflection, while $\CP$ symmetry corresponds to the vertical reflection (within the $0$-th order of $\alpha_s$).}
  \label{fig:DW}
\end{figure}

Now we review previous studies on the domain walls for the $\calN=1$ $\mathrm{SU}(N)$ SYM theory on small $\mathbb{R}^3\times S^1$, temporarily setting $m=0$. 
It has been found that the spatial Wilson loop shows the very peculiar behavior on the domain wall in confined vacua~\cite{Anber:2015kea, Sulejmanpasic:2016uwq, Cox:2019aji}. 
As we explained in Sec.~\ref{sec:other}, there are $N$ degenerate confined vacua due to $R$-symmetry breaking. Setting $\theta=0$ ($\theta$ is unphysical here), we see these vacua given by $M_i = M = \exp\left(2\pi \im k /N\right)$ with $k=0,1,\cdots,N-1$ [see Eq.~\eqref{eq:confined_vacua}]. A BPS domain wall between any two neighboring vacua, say $k=0$ and $k=1$, has $\mathbb{Z}_N$ distinct types, as such a domain wall should connect $M_i=1$ and $M_i=\exp(2\pi\im/N)$ continuously while satisfying $\prod_i M_i=1$, and those different types are related by $0$-form center symmetry. 
Now, let us put a fundamental Wilson loop on the domain wall which connects $k=0$ and $k=1$ vacua. 
Due to the presence of the Wilson loop on the domain wall, the BPS configurations inside and outside the Wilson loop should be different types because of the boundary condition. 
For example, when $N=2$, the domain walls inside and outside a Wilson loop, $W(C)$, along a contour $C$, should connect two vacua, $\sigma=0$ and $\sigma=\pi$ (which can be $\sim -\pi$ due to $\sigma\sim\sigma + 2\pi$), as 
\be
\sigma: \left\{\begin{array}{ll}
 0 \to +\pi  & (\mbox{outside of }C),\\[0.5em]
0\to -\pi & (\mbox{inside of }C). 
\end{array}\right.
\ee 
Still, these domain walls have the same wall tension, so the size of the Wilson loop does not affect its expectation value and we get the perimeter law~\cite{Anber:2015kea, Sulejmanpasic:2016uwq, Cox:2019aji}. 
This leads to the 1-form deconfinement on domain walls~\cite{Anber:2015kea}. 
A recent study in Ref.~\cite{Cox:2019aji} extended the analysis to non-neighboring vacua and explicitly confirmed the conjectured behavior of the domain wall given in Ref.~\cite{Acharya:2001dz}. We note that the deconfinement on the wall also occurs for the $\mathcal{CP}$ domain walls for the softly-broken $\calN=1$ SYM theory, and even for pure YM theories, at $\theta=\pi$. 
Indeed, this turns out to be a part of the requirement to match the 't~Hooft anomaly by spontaneous $\mathcal{CP}$ breaking~\cite{Gaiotto:2017yup}. 

We have found that the $\mathrm{SU}(2)$ SYM theory has a novel deconfined and $\mathcal{CP}$-broken phase, so it is interesting to look at the domain walls in this phase. 
The story about the Wilson line and domain walls turns out to be more interesting here. 
We number the vacua from 1 to 4 as $\sigma$ increases from $0$ to $2\pi$, as shown in Fig.~\ref{fig:DW}. 
The $0$-form center symmetry relates $4\leftrightarrow1$ and $2\leftrightarrow3$ while $\CP$ symmetry relates $1\leftrightarrow2$ and $3\leftrightarrow4$. We only have 4 kinds of domain walls between each pair of vacua with adjacent numbers (1 is next to 4). Other domain walls such as $1\leftrightarrow 3$ and $2\leftrightarrow 4$ could exist only if $\varphi$ becomes infinite on the wall, which invalidates the infrared effective theory here. We refer to the domain wall between $i$ and $j$ vacua as DW$_{ij}$. Figure~\ref{fig:DW} also illustrates these domain walls.

From  Fig.~\ref{fig:DW}, it is easy to see that DW$_{12}$ and DW$_{34}$ are $\CP$ symmetric and related by 0-form center symmetry. If we change $\gamma$ and a phase transition to the confined phase occurs, these two walls become the two vacua of the $\CP$ domain wall there. If $\gamma$ gets large enough to realize the $\CP$-symmetric deconfined phase, these walls disappear. Hence the tensions of these walls decrease as $\gamma$ increases. The discussion about DW$_{41}$ and DW$_{23}$ is just the reverse and thus the tension increases as $\gamma$ increases. Note that the tension of every domain wall is of $\calO(\alpha_s)$ here (compared to the domain wall tension in the confined phase). 

\begin{figure}[t]
  \includegraphics[width = 0.72 \columnwidth]{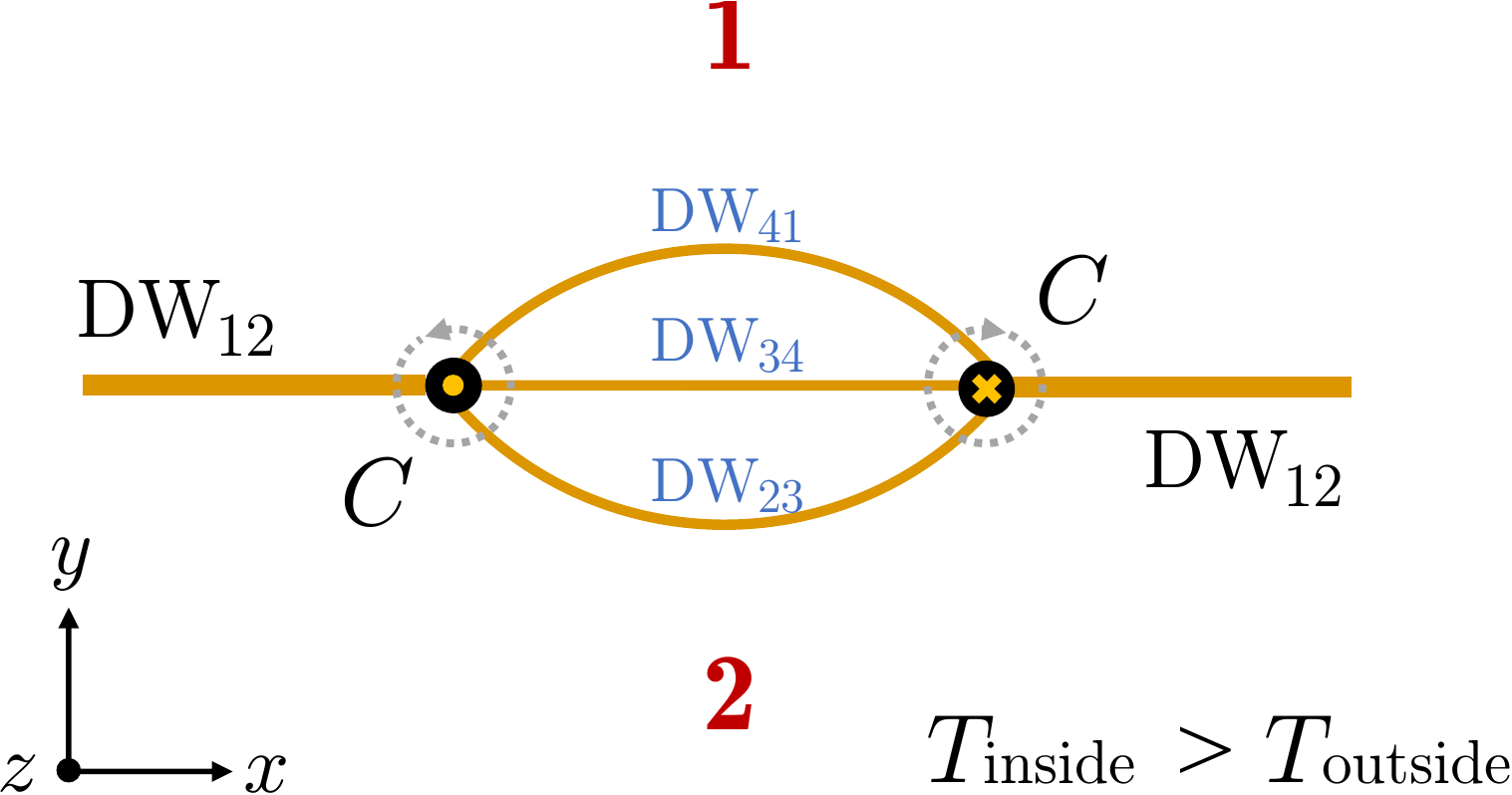}
  \caption{A transverse slice of a fundamental Wilson loop on the domain wall DW$_{12}$. The $z$ direction is perpendicular to the page and we only show the $x$-$y$ plane at $z=0$. DW$_{12}$ is located at $y=0$. In the $y>0$ region the vacuum is 1 and in the $y<0$ region the vacuum is 2. The black dots denote the $z=0$ section of a fundamental Wilson loop $C$ on DW$_{12}$. Due to the $2\pi$ monodromy of $\sigma$ indicated by the dashed circle arrows, inside the loop a sequence of DW$_{23}$, DW$_{34}$ and DW$_{41}$ appear. As a consequence, the wall tensions inside and outside the Wilson loop are different, and the Wilson loop on the wall obeys the are law.  }
  \label{fig:wilson}
\end{figure}

Now we try to put a spatial Wilson loop on a domain wall. For example, let us put a fundamental Wilson loop on DW$_{12}$. Because of the nontrivial monodromy of $\sigma$, inside the loop $C$, DW$_{12}$ is replaced by a sequence of DW$_{23}$, DW$_{34}$ and DW$_{41}$. This configuration is schematically illustrated in Fig.~\ref{fig:wilson}. Since DW$_{34}$ has the same tension as DW$_{12}$, the interior tension is larger than the exterior tension, namely,
\begin{equation}
	T_{\mathrm{inside}}>T_{\mathrm{outside}}.
\end{equation}
Indeed, we can approximately think of the interior tension as the sum of that of each individual wall\footnote{We note that this is a good approximation to describe the confining string of this model, at least in the confinement phase~\cite{Anber:2015kea}. In order to get an intuitive understanding, we here assume that its validity extends to the novel phase, but this is not an essential part of our discussion. }, i.e. $T_{\mathrm{inside}}\simeq\mathrm{T}(\mathrm{DW}_{23}+\mathrm{DW}_{34}+\mathrm{DW}_{41})$. It is surely larger than $T_{\mathrm{outside}}\equiv\mathrm{T}(\mathrm{DW}_{12})=\mathrm{T}(\mathrm{DW}_{34})$. Consequently, this Wilson loop on the wall, $\mathrm{DW}_{12}$, has an area law.
Similar arguments work for other domain walls. As a conclusion, the 1-form deconfinement on domain walls does not take place in the novel phase.

\begin{figure}[t]
  \includegraphics[width = 0.53\columnwidth]{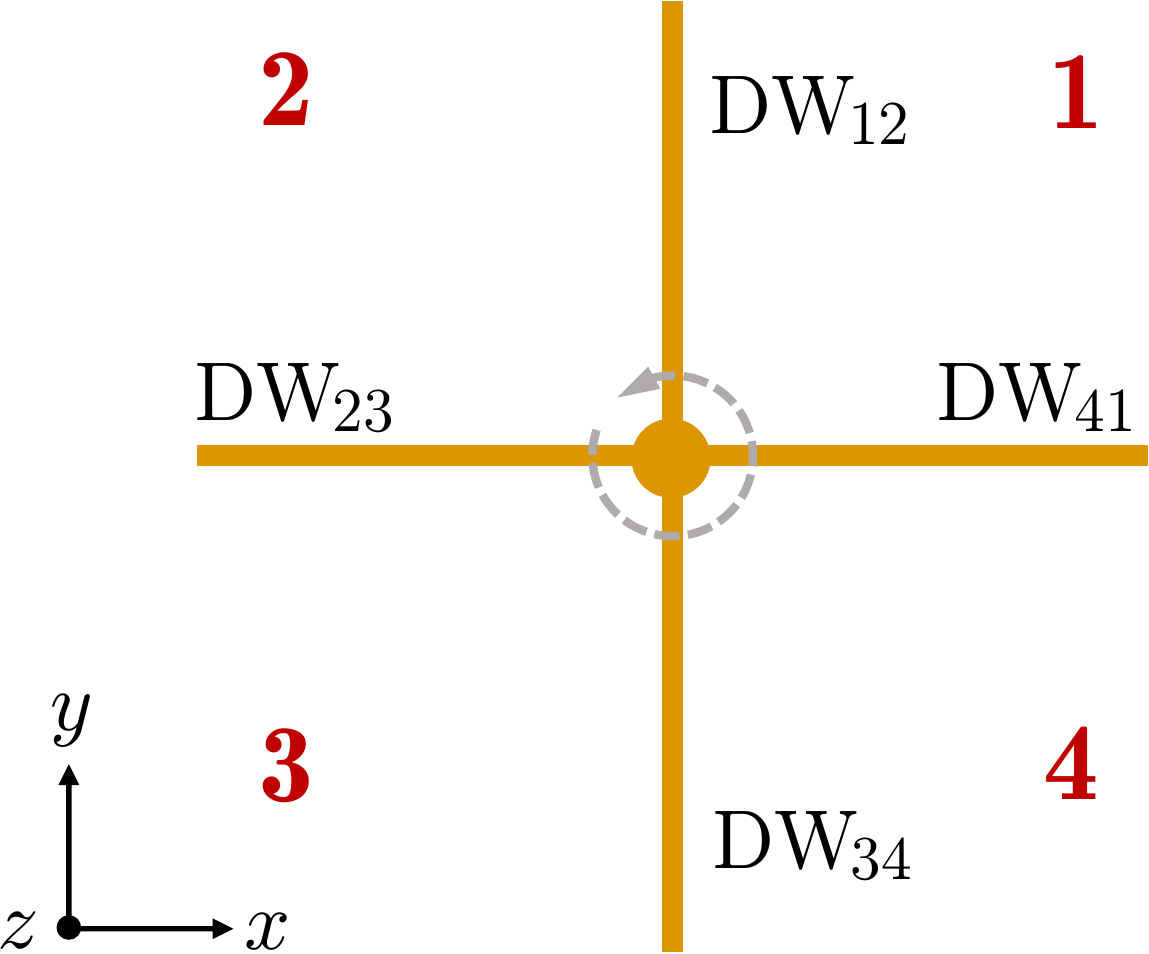}
  \caption{A transverse slice of a domain wall junction. The $z$ direction is perpendicular to the page and we only show the $x$-$y$ plane at $z=0$. The junction is located at $x=y=0$, namely along the $z$-axis. The dashed circle arrow indicates the $2\pi$ monodromy of $\sigma$ around the junction.}
  \label{fig:junction}
\end{figure}

According to the above discussion, the domain-wall junction must be accompanied by a nontrivial electric charge. With the list of domain walls at hand, the only possible junction is the one surrounded by DW$_{12}$, DW$_{23}$, DW$_{34}$, and DW$_{41}$ sequentially. This configuration is illustrated in Fig.~\ref{fig:junction}. In other words, $\sigma$ has a $2\pi$ monodromy around this junction. Thus, this junction is nothing but a fundamental Wilson line. That is to say, the domain wall junction can be present only when we introduce a nontrivial electric charge by the external Wilson line.\footnote{
We note that, although a similar gauge theory with four vacua has been considered in Ref.~\cite{Anber:2020xfk} by replacing the $\theta$ parameter to the dynamical axion field with massless fermions, the anomaly still requires the deconfinement on the wall. The non-deconfinement on the wall and the special nature of junctions are very unique to the novel phase of $\mathrm{SU}(2)$ gauge theory. }

The story becomes more interesting when we consider the $\mathrm{SU}(2)$ YM theory with the heavy but dynamical fundamental quarks, and we assume that their mass is much larger than the strong scale, $m_{\mathrm{fnd}}\gg \Lambda$. 
Most of our discussions are unaffected as the dynamical quarks are sufficiently heavy. Now, however, the domain-wall junction is allowed as the dynamical object, and then the fundamental quarks should be pinned to the location of the junction. 

All the phenomena above should be consistent with the 't~Hooft anomaly and play some roles in the anomaly matching. First, from Fig.~\ref{fig:DW} we see that either 0-form center symmetry or $\CP$ symmetry is not manifested on a domain wall. Hence the manifestation of the 1-form center symmetry on domain walls is not contradictory to the 't Hooft anomaly. Second, because the junction is simultaneously 0-form center symmetric and $\CP$ symmetric, as shown in Fig.~\ref{fig:junction}, to match the anomaly, the junction must be the charged object under the $1$-form center symmetry. Other scenarios are rejected by the dimension; the junction is only 1-dimensional and not capable to support other scenarios. In other words, the nontrivial electric charge of the junction is protected by the 't~Hooft anomaly of the bulk theory. As a conclusion, we have demonstrated that the 't Hooft anomaly requires novel properties on the domain walls and the junction in the phase we discovered.

\section{summary}\label{sec:summary}

In this work, we studied the realization of the phase structure constrained by the 't Hooft anomaly between center and $\CP$ symmetries in the thermal YM theory and the softly-broken $\calN=1$ SYM theory. We quantitatively evaluated the phase diagram from the infrared effective potential in the latter theory on small $\mathbb{R}^3\times S^1$, with the hope of optimistic extrapolation of the qualitative behavior to the former theory.

This work exemplified the 't Hooft anomaly matching argument in a concrete model and also went further favoring a particular scenario among many possibilities. 
For all the gauge groups $G$ but $\mathrm{SU}(2)$, we confirmed that the deconfinement transition is of the first order. Also, at $\theta=\pi$, the deconfinement and the $\mathcal{CP}$ restoration occur at the same temperatures, and we expect that this is most likely true also for the thermal YM theories. 

Interestingly, $\mathrm{SU}(2)$ gauge group turns out to be very special. 
At $\theta=\pi$, a novel deconfined phase with a broken 0-form center symmetry as well as a broken $\CP$ symmetry was unraveled for $G=\mathrm{SU}(2)$, which has a sharp contrast with other gauge groups for which the phase transition of simultaneous deconfinement and $\CP$-restoration is observed. The SU(2) phase diagram is accompanied by a rare intersection between a first-order and a second-order phase transitions. We also illustrated $1$-form center symmetry in this novel phase, especially showing that the domain-wall junction must be a charged object of the $1$-form symmetry. 
All of these features find their ways to match the 't~Hooft anomaly.

Although our discoveries are conclusive within the softly-broken $\calN=1$ SYM theory on small $\mathbb{R}^3\times S^1$, the extrapolation to the thermal YM theory should be more circumspect. 
The phase structure around $\theta=\pi$ of the pure $\mathrm{SU}(2)$ gauge theory might be even more exotic than our results. This question in the thermal YM theory deserves further investigations.

\begin{acknowledgments}
The authors thank 
Mohamed~Anber, Erich~Poppitz, and Mithat~\"{U}nsal
for useful discussions and also for useful comments on early drafts.
The authors especially thank Robert~D.~Pisarski for having an interesting conversation, which really gave us the motivation to initiate this study. 
K.~F.\ was partially supported by Japan Society for the Promotion of Science
(JSPS) KAKENHI Grant Nos.\ 18H01211 and 19K21874.
\end{acknowledgments}

\appendix 
\section{Simultaneous Deconfinement and\\ $\mathcal{CP}$ Restoration for non-$\mathrm{SU}(N)$ gauge groups}
\label{app:nonSU}

In this Appendix, we extend our discussion to non-$\mathrm{SU}(N)$ gauge groups, and confirm that deconfinement and $\mathcal{CP}$ restoration happen at the same temperature $\gamma$. 
We note that the effective potential (\ref{eq:potential}) should be replaced by a more complicated form, which can be found in Ref.~\cite{Anber:2014lba}.
 And also the center symmetry ceases to be generated by \eqref{eq:symmetry3}, but is still a subgroup of the symmetric group of $\{M_i\}$. Fortunately, $\CP$ symmetry remains the same as \eqref{eq:symmetry3}. 

We would like to briefly comment on the anomaly in Sec.~\ref{sec:pureYM} for non-$\mathrm{SU}(N)$ gauge groups. 
In the presence of this $\mathcal{CP}$-center mixed anomaly, $Q$ has to be fractionalized by the background $2$-form gauge field $B$. 
Let us give a list of the simply-connected gauge groups, with its center and the minimal fractionalized value of $Q$~\cite{Witten:2000nv, Aharony:2013hda}:
\be
\begin{array}{c|c|c}
G & \mathcal{Z}(G) & Q[B] \bmod 1 \\ \hline
\mathrm{SU}(N) & \mathbb{Z}_N & 1/N \\ 
\mathrm{Spin}(2k+1) & \mathbb{Z}_2 & 0\\
\mathrm{Spin}(4k) & \mathbb{Z}_2\times \mathbb{Z}_2 & 1/2\\
\mathrm{Spin}(4k+2) & \mathbb{Z}_4 & 1/4\\
\mathrm{Sp}(2k) & \mathbb{Z}_2 & 0\\
\mathrm{Sp}(2k+1) & \mathbb{Z}_2 & 1/2\\
E_6 & \mathbb{Z}_3 & 2/3 \\
E_7 & \mathbb{Z}_2 & 1/2\\
E_8,\; F_4,\; G_2 & 1 & 0
\end{array}
\ee
In this table, $\mathrm{Spin}(N)$ is always understood to be $N\ge 5$. 
For gauge groups with $Q$ mod $1$ being $0$, the mixed anomaly is not present even in the generalized sense of Refs.~\cite{Kikuchi:2017pcp, Karasik:2019bxn}. Thus, the constraint discussed above does not exists for those gauge groups, $\mathrm{Spin}(2k+1)$, $\mathrm{Sp}(2k)$, and exceptional groups $E_8, F_4, G_2$ without center.

\begin{table}
  \centering
\begin{tabular}{|c|c|l|c|}
     \hline $G$ & $\mathcal{Z}(G)$ & \multicolumn{1}{c|}{$M_{i}$} & $\gammadec(\pi)$ \\\hline\hline 
     
     $\mathrm{Spin}(8)$ & $\mathbb{Z}_2$$\times $$\mathbb{Z}_2$ & \begin{tabular}{l}
          -0.019, 2.383, 1.984, 2.383,\\ 2.383
     \end{tabular} & 1.521 \\\hline 
     
     $\mathrm{Spin}(10)$ & $\mathbb{Z}_4$ & \begin{tabular}{l}
          -0.005, 2.015, 1.762, 2.266,\\
          2.518, 2.518
     \end{tabular} & 0.991 \\\hline 
     
     $\mathrm{Spin}(12)$ & $\mathbb{Z}_2$$\times $$\mathbb{Z}_2$ & \begin{tabular}{l}
          -0.001, 1.742, 1.567, 2.090,\\ 2.438, 2.612, 2.612
     \end{tabular} & 0.693 \\\hline 
     
     $\mathrm{Spin}(14)$ & $\mathbb{Z}_4$ & \begin{tabular}{l}
         -4$\times$$10^{\text{-4}}$, 1.532, 1.404, 1.915,\\ 2.298, 2.553, 2.681, 2.681
     \end{tabular} & 0.509 \\\hline 
     
     $\mathrm{Spin}(16)$ & $\mathbb{Z}_2$$\times $$\mathbb{Z}_2$ & \begin{tabular}{l}
          -1$\times$$10^{\text{-4}}$, 1.366, 1.268, 1.756,\\ 2.146, 2.439, 2.634, 2.732,\\ 2.732
     \end{tabular} & 0.390 \\\hline 
     
     $\mathrm{Spin}(18)$ & $\mathbb{Z}_4$ & \begin{tabular}{l}
          -3$\times$$10^{\text{-5}}$, 1.231, 1.154, 1.616,\\ 2.001, 2.308, 2.539, 2.693,\\
          2.770, 2.770
     \end{tabular} & 0.308 \\\hline 
     
     $\mathrm{Sp}(3)$ & $\mathbb{Z}_2$ & \begin{tabular}{l}
          -0.082, 1.535, 2.579, 3.095
     \end{tabular} & 1.847 \\\hline 
     
     $\mathrm{Sp}(5)$ & $\mathbb{Z}_2$ & \begin{tabular}{l}
          -0.024, 1.028, 1.856, 2.473,\\ 2.883, 3.088
     \end{tabular} & 0.780 \\\hline 
     
     $\mathrm{Sp}(7)$ & $\mathbb{Z}_2$ & \begin{tabular}{l}
          -0.008, 0.766, 1.423, 1.970,\\ 2.406, 2.733, 2.951, 3.060
     \end{tabular} & 0.426 \\\hline 
     
     $\mathrm{Sp}(9)$ & $\mathbb{Z}_{2}$ & \begin{tabular}{l}
          -0.003, 0.610, 1.152, 1.626,\\ 2.032, 2.370, 2.640, 2.843,\\
          2.978, 3.045
     \end{tabular} & 0.268 \\\hline 
     
     $\mathrm{E}_6$ & $\mathbb{Z}_{3}$ & \begin{tabular}{l}
          -3$\times$$10^{\text{-4}}$, 2.382, 2.233, 2.084,\\ 2.233, 2.382, 1.638
     \end{tabular} & 0.595 \\\hline 
     
     $\mathrm{E}_7$ & $\mathbb{Z}_{2}$ & \begin{tabular}{l}
          -4$\times$$10^{\text{-6}}$, 1.511, 1.955, 2.133,\\ 2.221, 2.310, 2.399, 2.177
     \end{tabular} & 0.355 \\\hline 
\end{tabular}
  \caption{Numerically evaluated $\gammadec(\pi)=\gammaCP$ for the gauge groups with a mixed 't Hooft anomaly, including $\mathrm{Spin}(2k)$, $\mathrm{Sp}(2k+1)$, $\mathrm{E}_6$, and $\mathrm{E_7}$. 
  }
  \label{tab:B}
\end{table}

In Table~\ref{tab:B}, we list the numerical computation for gauge groups with a mixed 't Hooft anomaly between center and $\CP$ symmetries.
The qualitative features discussed in $\mathrm{SU}(N)$ turn out to be correct also for these gauge groups. Namely, one negative suppressed amplitude is followed by positive magnified ones. The $\mathbb{Z}_2$ $\mathcal{C}$-symmetry is also unbroken for $\mathrm{Spin}(2k)$ with $k>4$ as well as $\mathrm{E}_6$, so some of their $M_i$ appear in pairs. Since the $\mathcal{C}$-symmetry of $\mathrm{Spin}(8)$ is $S_3$, we see three identical amplitudes. Other groups in Table~\ref{tab:B} has no charge conjugation. Most importantly, the imaginary parts of $M_i$ in Table~\ref{tab:B} are also zero up to $10^{-10}$, which implies that $\gammadec(\pi)=\gammaCP$ holds for these groups as well.

\begin{table}
  \centering
\begin{tabular}{|c|c|l|c|}
     \hline $G$ & $\mathcal{Z}(G)$ & \multicolumn{1}{c|}{$M_{i}$} & $\gammadec(\pi)$ \\\hline\hline 
     $\mathrm{Spin}(5)$ & $\mathbb{Z}_2$ & \begin{tabular}{l}
          -0.170, 1.966, 2.990
     \end{tabular} & 3.416 \\\hline 
     
     $\mathrm{Spin}(7)$ & $\mathbb{Z}_2$ & \begin{tabular}{l}
          -0.037, 2.610, 2.085, 2.350
     \end{tabular} & 1.950 \\\hline 
     
     $\mathrm{Spin}(9)$ & $\mathbb{Z}_2$ & \begin{tabular}{l}
          -0.010, 2.185, 1.871, 2.340,\\ 2.496
     \end{tabular} & 1.215 \\\hline 
     
     $\mathrm{Spin}(11)$ & $\mathbb{Z}_2$ & \begin{tabular}{l}
          -0.003, 1.869, 1.661, 2.180,\\ 2.491, 2.594
     \end{tabular} & 0.822 \\\hline 
     
     $\mathrm{Spin}(13)$ & $\mathbb{Z}_2$ & \begin{tabular}{l}
          -7$\times$$10^{\text{-4}}$, 1.631, 1.482, 2.001,\\ 2.371, 2.593, 2.668
     \end{tabular} & 0.591 \\\hline 
     
     $\mathrm{Spin}(15)$ & $\mathbb{Z}_2$ & \begin{tabular}{l}
          -2$\times$$10^{\text{-4}}$, 1.445, 1.333, 1.833,\\ 2.222, 2.500, 2.667, 2.722
     \end{tabular} & 0.444 \\\hline 
     
     $\mathrm{Spin}(17)$ & $\mathbb{Z}_2$ & \begin{tabular}{l}
          -6$\times$$10^{\text{-5}}$, 1.295, 1.209, 1.684,\\ 2.072, 2.374, 2.590, 2.720,\\ 2.763
     \end{tabular} & 0.345 \\\hline 
     
     $\mathrm{Spin}(19)$ & $\mathbb{Z}_2$ & \begin{tabular}{l}
          -2$\times$$10^{\text{-5}}$, 1.173, 1.104, 1.552,\\ 1.932, 2.242, 2.484, 2.656,\\ 2.760, 2.794
     \end{tabular} & 0.276 \\\hline 
     
     $\mathrm{Sp}(4)$ & $\mathbb{Z}_2$ & \begin{tabular}{l}
          -0.043, 1.236, 2.174, 2.793,\\ 3.102
     \end{tabular} & 1.148 \\\hline 
     
     $\mathrm{Sp}(6)$ & $\mathbb{Z}_2$ & \begin{tabular}{l}
          -0.014, 0.878, 1.613, 2.198,\\ 2.635, 2.927, 3.072
     \end{tabular} & 0.564 \\\hline 
     
     $\mathrm{Sp}(8)$ & $\mathbb{Z}_2$ & \begin{tabular}{l}
          -0.005, 0.679, 1.273, 1.782,\\ 2.205, 2.544, 2.797, 2.966,\\ 3.051
     \end{tabular} & 0.333 \\\hline 
     
     $\mathrm{E}_8$ & 1 & \begin{tabular}{l}
          -1$\times$$10^{\text{-9}}$, 2.214, 2.190, 2.165,\\ 2.117, 2.021, 1.829, 1.396,\\ 2.182
     \end{tabular} & 0.192 \\\hline
     
     $\mathrm{F}_4$ & 1 & \begin{tabular}{l}
          -0.002, 1.755, 2.194, 2.304,\\ 2.413
     \end{tabular} & 0.870 \\\hline
     
     $\mathrm{G}_2$ & 1 & \begin{tabular}{l}
          -0.076, 2.275, 2.534
     \end{tabular} & 2.742 \\\hline 
\end{tabular}
  \caption{Numerically evaluated $\gammadec(\pi)=\gammaCP$ for the gauge groups without a mixed 't Hooft anomaly, including $\mathrm{Spin}(2k+1)$, $\mathrm{Sp}(2k)$, $\mathrm{E}_8$, $\mathrm{F}_4$, and $\mathrm{G}_2$. }
  \label{tab:C}
\end{table}

We list in Table~\ref{tab:C} the numerical results for gauge groups without a mixed 't Hooft anomaly between center and $\CP$ symmetries. All the features look familiar as previous tables. This time none in Table~\ref{tab:C} has a charge conjugation. The imaginary parts of $M_i$ again vanish up to $10^{-10}$. In these cases, the phase structure is not ordered by the 't~Hooft anomaly. Nevertheless, the confining vacuum still has a multi-branch structure characterized by the dual Coxeter number $c_2$, and the theories show the spontaneous $\mathcal{CP}$ breaking at $\theta=\pi$ in the confined phase. 
It is quite interesting to see that $\gammadec(\pi)=\gammaCP$ still holds even when 't~Hooft anomaly does not require it. 

More explanations are needed for $\mathrm{E}_8$, $\mathrm{F}_4$, and $\mathrm{G}_2$ in the table, whose centers are trivial. 
Generally speaking, the generalization of the potential \eqref{eq:potential} leads to a deconfinement-like phase transition for them~\cite{Poppitz:2012nz,Anber:2014lba}. 
However, this phase transition cannot be characterized by the Landau criterion using the center symmetry. 
It is merely a $\bphi$ jump instead, which would be caused by the huge difference between the numbers of confined degrees of freedom and of deconfined gluons. 
The dimensions of the Lie groups with trivial center,  $\mathrm{E}_8$, $\mathrm{F}_4$, and $\mathrm{G}_2$, are $248$, $52$, and $14$, which roughly correspond to $N\simeq 50$, $7$, and $4$ for $\mathrm{SU}(N)$ gauge group, respectively. These numbers of gluons may be large enough to compare the confinement-deconfinement transition with the large-$N$ Hagedorn-type first-order transition.

Now we complete all the numerical results for gauge groups whose rank is less than 10. Although we have not checked the whole gauge groups, we are tempted to conclude that we always have $\gammadec(\pi)=\gammaCP$ for all $G\not=\mathrm{SU}(2)$. 
As we have discussed in the main part of this paper, we expect that this equality might be extrapolated to the pure thermal Yang-Mills theory, i.e., $\Tdec(\pi)=\Tcp$ for all $G\not=\mathrm{SU}(2)$.

\section{Deconfinement transition in large $N$}
\label{app:largeN}

In this section, we discuss the confinement-deconfinement transition of the effective potential (\ref{eq:potential}) in the large-$N$ limit. We shall find the analytic expression for the deconfinement temperatures $\gammadec(\theta)$ and also the expectation values of $M_i$. 
As we have done in our numerical analysis in Sec.~\ref{sec:other}, we neglect the $\mathcal{O}(\alpha_s)$ correction, and then the effective potential (\ref{eq:potential}) can be written as 
\be
{\mathcal{V}\over V_0}=\sum_i |M_i-M_{i-1}|^2-{\gamma\over 2}\sum_i (M_i+M_i^*). 
\label{eq:potential_app}
\ee

Since we already know about the confinement phase (\ref{eq:confined_vacua}), our interest is to obtain the deconfined vacua. 
For this purpose, let us introduce 
\be
X(t_i)=M_{N/2+N t_i}, 
\ee
with 
\be
t_i=-{1\over 2}+{i\over N}\quad (i=0,1,\ldots, N-1). 
\ee 
This implies the usefulness of the continuum approximation, such as $M_i-M_{i-1}\simeq {1\over N}\partial_t X(t)$. We obtain that 
\be
{\mathcal{V}\over V_0}={1\over N}\int_{-1/2}^{1/2}\diff t \left(|\p_t X|^2-{N^2 \gamma \over 2} (X+X^*)\right). 
\label{eq:potential_cont}
\ee

So far, we have neglected the constraints (\ref{eq:constraint_m}). 
In order to take it into account, we put an extra insight suggested by the numerical analysis in Sec.~\ref{sec:other}: In the deconfined vacua, most of the monopole operators get expectation values of $\mathcal{O}(1)$, and the one monopole has quite small expectation value.  
Indeed, when $M_{i\not=0}\sim 1$, then $\prod_{i\not=0}M_i$ typically becomes an exponentially large or small number unless fine tuned. 
Assuming that $\prod_{i\not=0}M_i$ turns out to be exponentially large, then $M_0$ has to be exponentially small in $N$ in order to satisfy the constraint (\ref{eq:constraint_m}). 
Therefore, we can approximately set $M_0=0$ in the large-$N$ limit when analyzing the effective potential. 
This means that we have to treat $|M_1-M_0|^2$ and $|M_{0}-M_{N-1}|^2$ in (\ref{eq:potential_app}) separately, which gives additional terms to the continuum expression (\ref{eq:potential_cont}). 
We obtain 
\bea
{\mathcal{V}\over V_0}&=&{1\over N}\int_{-1/2}^{1/2}\diff t \left(|\p_t X|^2-{N^2 \gamma \over 2} (X+X^*)\right)\nonumber\\
&&+|X(-1/2)|^2+|X(1/2)|^2. 
\label{eq:potential_cont2}
\eea
The equation of motion is given by 
\be
\p_t^2 X = -{N^2\gamma\over 2}.  
\ee
Since the boundary term gives the $\mathcal{O}(1)$ contribution while the others are only of $\mathcal{O}(1/N)$, we obtain the Dirichlet boundary condition, 
\be
X(-1/2)=X(1/2)=0. 
\ee
The solution is 
\be
X(t)={N^2\gamma\over 16}(1-4t^2). 
\ee
Substituting this expression into (\ref{eq:potential_cont2}), we obtain the free energy for the deconfined phase, 
\be
{\mathcal{V}_{\mathrm{dec}}\over V_0}=-{(N^2\gamma)^2\over 48 N}. 
\ee
Especially, since the continuum approximation of the potential (\ref{eq:potential_cont2}) does not have the $\theta$ dependence, the free energy of the deconfined phase is independent of $\theta$. 
The whole $\theta$ dependence is carried by $M_0$, but it is exponentially small in $N$. 
The rough estimate indeed shows that 
\be
M_0\simeq {1\over N}\left({4e^2\over N^2\gamma}\right)^{N-1}\exp\left(\im \theta\right),  
\ee
and thus our ansatz turns out to be self-consistent as long as $N^2\gamma > 4e^2\simeq 30$ (here $e$ is the base of natural logarithm). 

In order to evaluate the deconfinement temperature $\gammadec(\theta)$, we compare the free energy $\mathcal{V}_{\mathrm{dec}}$ with that of the confined phase:
\be
{\mathcal{V}_{\mathrm{conf}}\over V_0}=-N\gamma \cos\left({\theta\over N}\right). 
\ee
We therefore obtain that 
\be
\gammadec(\theta)={48\over N^2}\left(1-{\theta^2\over 2N^2}\right)
\ee
for $-\pi<\theta<\pi$. Since $N^2\gamma\simeq 48> 4e^2$, this is in the valid range of our ansatz. 
Moreover, if we take $N=10$ at $\theta=\pi$ as an example, then this formula predicts $\gammadec(\pi)\simeq 0.456$. This shows a good agreement with Table~\ref{tab:A}. 

We can also evaluate the expectation values of the monopole operators just above the deconfinement temperature. 
For BPS monopoles, they are given by 
\be
M_{i}=X(t_i)=3\left(1-{\theta^2\over 2N^2}\right) (1-4t_i^2), 
\ee
for $t_i=-1/2+i/N$ with $i\not = 0$. 
For KK monopole,  
\be
M_0={1\over N}\left({e^2\over 12}\right)^{N-1}\left(1+{\theta^2\over 2N}\right)\exp\left(\im \theta\right).
\ee 
These values also roughly agree with those of Table~\ref{tab:A}. Especially, this result shows that $\mathcal{CP}$ restoration occurs at $\gammadec(\pi)$ in the large-$N$ limit.

\bibliography{./novel}
\bibliographystyle{utphys.bst}
\end{document}